\documentclass[%
aps,
prd,
twocolumn,
preprintnumbers,
nofootinbib,
amsmath,amssymb,
]{revtex4-2}

\usepackage{style}
\usepackage{aas_macros}

\begin{document}

\preprint{DESY-24-063}

\title{Astrometric Search for Ultralight Dark Matter}

\author{Hyungjin Kim\,\orcidlink{0000-0002-8843-7690}}
\affiliation{Deutsches Elektronen-Synchrotron DESY, Notkestr. 85, 22607 Hamburg, Germany}

\begin{abstract}
Precision astrometry offers a way to probe new physics. By measuring the angular position of light sources at unprecedented precision, astrometry could probe minuscule fluctuations of underlying spacetime. This work explores the possibility of probing ultralight dark matter candidates using precision astrometry. Through the coherent and stochastic density fluctuations over the scale of its wavelength, ultralight dark matter perturbs the propagation of light and the geodesics of the observer and source, leading to unique time-dependent signatures in the angular position of background light sources. With detector specifications similar to the current and future astrometry observations, such as Gaia and Roman Space Telescope, it is shown that the ultralight scalar dark matter of mass $10^{-18}\eV \, \textrm{--} \, 10^{-16} \eV$ could be probed when its density near the solar system is about a few thousand times larger than the nominal dark matter density measured on a much larger kpc-scale. This sensitivity is comparable to current pulsar timing array observations at a similar mass range. Explicit expressions for the angular deflection induced by most generic metric perturbations are derived and its gauge invariance is explicitly checked at the linear order. 
\end{abstract}

\maketitle
\tableofcontents

\section{Introduction}
Precision astrometry measures the positions of astrophysical objects at unprecedented precision. The current Gaia mission provides precise astrometric solutions for a billion of light sources~\cite{2023A&A...674A...1G}, while the $40$-year Very Long Baseline Interferometry observations of quasi-stellar objects provides an absolute celestial reference frame at the precision of ${\rm \mu as}= 5\times 10^{-12}\,{\rm rad}$~\cite{2020A&A...644A.159C}. Due to their precision, these measurements could provide an interesting way to probe gravity waves and new physics. As the fluctuations of underlying spacetime perturb the angular position of light sources, any physics scenario predicting significant spacetime fluctuations in the present universe could leave distinctive signatures in those measurements. These signatures could then be used to investigate the gravitational wave background~\cite{1990NCimB.105.1141B, 1996ApJ...465..566P, Book:2010pf, Raccanelli:2016fmc, Klioner:2017asb, Moore:2017ity, Wang:2020pmf, Wang:2022sxn, Caliskan:2023cqm}, and beyond-the-standard model scenarios, such as those with compact dark objects and dark matter substructures~\cite{Li:2012qha, VanTilburg:2018ykj, Mondino:2020rkn, Mishra-Sharma:2020ynk, Mondino:2023pnc, DeRocco:2023gde, Fardeen:2023euf, DeRocco:2023hij}. 

Ultralight dark matter (ULDM) constitutes an interesting alternative dark matter candidate. It is defined as a bosonic dark matter with a mass smaller than several eV scales. Being bosonic and light, it behaves similarly to classical waves rather than particles,  resembling classical electromagnetic waves but with the dispersion relation of massive particles. One of the interesting implications of the wave nature of these DM candidates is that they exhibit order one density fluctuations over the astrophysical scale. These density fluctuations have been observed through numerical simulations (e.g.~\cite{Schive:2014dra}). Various phenomenological consequences have been studied especially in conjunction with observed cold stellar substructures~\cite{Marsh:2018zyw, Amorisco:2018dcn, Dalal:2020mjw, Dalal:2022rmp}, constraining the existence of such dark matter candidates with its mass $m \lesssim 10^{-19}\eV$. 

What is even more interesting is that the density fluctuation comes with two distinctive temporal and spectral behaviors. In particular, the density fluctuation can be decomposed into the fast and slow modes as~\cite{Kim:2023pkx}
$$
\delta \rho(t,\boldsymbol x)
= 
\delta \rho_F(t, \boldsymbol x)
+ \delta \rho_S(t, \boldsymbol x). 
$$
Here $\delta \rho_F(t, \boldsymbol x)$ represents the fast mode that oscillates at the frequency of $\omega_F =2m$ and $\delta \rho_S(t,\boldsymbol x)$ represents the slow mode with its characteristic frequency $\omega_S = m \sigma_v^2$ with a typical dark matter velocity dispersion $\sigma_v \simeq 160\,{\rm km/sec}$ in our Galaxy. In the frequency space, the spectrum of both modes has a width $\Delta \omega = m \sigma_v^2$. The simultaneous appearance of the fast and slow modes is because the density is quadratic in the ultralight dark matter field $\delta\rho \propto \phi^2$. What we refer to as an order one density fluctuation is represented by the slow mode fluctuation $\delta\rho_S(t,\boldsymbol x)$. 

The existence of two distinctive oscillating modes offers great opportunities for ULDM searches. One intriguing possibility is to use gravity wave detectors. While they are originally designed to measure minuscule fluctuations of spacetime induced by gravity waves, they could also probe the fluctuations of underlying spacetime induced by ultralight dark matter. With time series data of these detectors, one could closely examine either or both the fast and slow mode of ULDM fluctuations, depending on the frequency band of the detector under consideration. Previous studies along this line have examined the implications of the fast mode of density fluctuations in interferometric gravity wave detectors~\cite{Aoki:2016kwl, Kim:2023pkx, Yu:2024enm} and pulsar timing array~\cite{Khmelnitsky:2013lxt, Porayko:2018sfa, NANOGrav:2023hvm, EuropeanPulsarTimingArray:2023egv}, and of the slow mode in interferometric detectors~\cite{Kim:2023pkx} and in pulsar timing array~\cite{Kim:2023kyy}.

This work explores the possibility of probing ultralight dark matter candidates with precision astrometry. This work concerns both the fast and slow modes and focuses on assessing the time-dependent behavior of the density fluctuation in astrometric observations. In other words, we consider the characteristic time scale of fluctuation to be smaller than the observational period of a given mission, $T_{\rm obs} > 1/ \omega_{F, S}$, such that time-dependent ULDM signal is fully reflected in the angular deflection. The main objective is to characterize the signal power spectrum of the fast and slow mode of ULDM density fluctuations in the angular deflection and to project the sensitivities of the current and future astrometric observations. We only consider the gravitational interaction in this work.

The paper is organized as follows. In Section~\ref{sec:deflection}, the relevant expressions for the angular deflection are summarized. We estimate the angular deflection from the fast and slow modes of ULDM density fluctuations and present the expressions for the signal spectra and the correlation. In Section~\ref{sec:snr}, we estimate the sensitivity of the current and future missions based on the optimal cross-correlation statistic. Using two possible detector specifications that resemble the Gaia and Roman Space Telescope, we project the sensitivity of astrometric observations for ULDM searches near the solar system. Section~\ref{sec:discussion} examines several assumptions made in the main text. Concluding remarks and possible future directions are presented in Section~\ref{sec:conclusion}. In the Appendix, we discuss in detail the statistical properties of the energy-momentum tensor of ultralight dark matter, an expression for the angular deflection in the presence of most generic metric perturbations in the non-expanding universe, the vector spherical harmonics expansion of ULDM signals, and details of numerical simulation and optimal statistics discussed in the main text. Throughout this work, we use the natural unit  $c=\hbar =1$ and the mostly positive metric signature $\eta_{\mu\nu} = (- + + +)$.

\section{Angular Deflection}\label{sec:deflection}

Consider the photon four-momentum in the flat spacetime,
$$
k^\mu = \omega_0 ( 1 , - n^i) . 
$$
The photon propagates in the direction of $-n^i$ in the coordinate space. The photon four-momentum can also be written in the tetrad basis of an observer,
\begin{align}
k^\mu = 
\omega_o e^{\mu}_{\; 0} 
- \omega_o n^i_o e^{\mu}_{\; i} ,
\end{align}
where the vierbein $e^\mu_{\; a}$ erects an orthonormal coordinate system for the observer. The basis is normalized as $\eta_{ab} = g_{\mu\nu} e^{\mu}_{\; a} e^\nu_{\; b}$, and the zeroth component is aligned with the four-velocity of the observer, $e^\mu_{\; 0} = u^\mu$. The photon frequency measured by the observer is $\omega_o = - g_{\mu\nu} k^\mu u^\nu$. The photon direction measured by the proper frame of the observer is given by
\begin{align}
n^i_o = \frac{g_{\mu\nu} k^\mu e^\nu_{\; i}}{g_{\mu\nu} k^\mu e^\nu_{\; 0}}. 
\end{align}
The right-hand side is computed along the observer's worldline. We define the angular deflection as
\begin{align}
\delta n^i = n^i_o - n_s^i , 
\end{align}
where $-n_s^i$ is the photon propagation direction measured in the source frame. The subscript $o$ ($s$) denotes quantities evaluated at the time of observation (emission) at the observer (source) location. 

In curved spacetime, the photon four-momentum and the tetrad coordinate of the observer evolve non-trivially along their respective worldlines. The resulting angular deflection reflects the underlying metric fluctuations. Photon four-momentum and tetrad basis are parallel transported along their  worldlines,
\begin{align}
\frac{dk^\mu}{d\lambda} &= - \Gamma^\mu_{\;\;\nu\rho} k^\nu k^\rho , 
\\
\frac{d e^\mu_{\; a}}{d\tau} &= - \Gamma^\mu_{\;\;\nu\rho} e^\nu_{\; a} e^\rho_{\; 0},
\end{align}
where $\lambda$ and $\tau$ are the affine parameter of the photon and the proper time of the observer, respectively, and $\Gamma^\mu_{\;\;\nu\rho}$ is the Christoffel symbol. At the linear order in metric perturbations, we expand
\begin{align}
g_{\mu \nu} &= \eta_{\mu\nu} + \delta g_{\mu\nu},
\\
k^\mu &= k_0^\mu + \delta k^\mu,
\\
e^\mu_{\; a} &= \delta^\mu_{\; a} + \delta e^{\mu}_{\; a}.
\end{align}
Note that we expand the tetrad basis with respect to the one for the stationary observer; we treat the velocity of the observer as a small perturbation. With this, the angular deflection may be written as
\begin{align}
\delta n^i = 
[\delta n^i]_{\rm L}
+ [\delta n^i]_{\rm A}.
\label{deflection_general}
\end{align}
We call $[\delta n^i]_{\rm L}$ and $[\delta n^i]_{\rm A}$ as a lensing and an aberration term, respectively. The detailed expression for each term is given below.

\bigskip 

\noindent\textbullet~\emph{Lensing term:} The first term is given by
\begin{align}
[\delta n^i]_{\rm L}
&= - \frac{P^{ij}(n)}{2\omega_0} \int^{\lambda_o}_{\lambda_s} d\lambda' \, 
\delta g_{\mu\nu,j} (\lambda') k_0^\mu k^\nu_0 , 
\end{align}
where $P^{ij}(n) = \delta^{ij} - n^i n^j$ is a projection tensor, and $\lambda_{o,s}$ is the affine parameter of a photon at the event of observation and emission, respectively. We call this term \textit{a lensing term} as it is due to the metric fluctuation in between the source and the observer.

\medskip

\noindent\textbullet~\emph{Aberration term:} The second term is given by
\begin{align}
[\delta n^i]_{\rm A}
&= - \frac{ P^{ij}(n) }{ \omega_0 } 
\eta_{\mu\nu} k^\mu_0 
\Big[
(\delta e^\nu_{\; j})_{o}
- (\delta e^\nu_{\; j})_{s}
\Big]. 
\label{deflection_aberration}
\end{align}
This term is due to the fluctuation of the reference frame of the observer and the source. In the absence of the metric perturbation but with the relative motion between the observer and the source, the above gives rise to $\delta n^i = P^{ij}(n) v_{\rm rel}^j$ with $v_{\rm rel}$ being the relative velocity between the observer and the source; this is a classical expression for the light aberration due to the relative motion between the observer and emitter, and hence, we name it an {\it aberration term}. 

\bigskip 

The relative importance of each term depends on the characteristics of the underlying spacetime fluctuations. For the angular deflection caused by the fast mode of the density fluctuations, the lensing and aberration terms are comparable. In contrast, for the slow mode of density fluctuations, the aberration term is dominant.

\subsection{Spectrum}
The ultralight dark matter signal can be characterized by its spectrum and the overlap reduction function, which describes the correlation of the angular deflection from different light sources. The main objective of this section is to summarize the signal power spectrum and the overlap reduction function resulting from ULDM density fluctuations. 

The covariance matrix of the angular deflection is given by
\begin{align}
\langle
\delta n_a^i(t) \delta n_b^j(t') 
\rangle
= \int_{f_l}^{f_u} df \, S(f) \Gamma_{ab}^{ij} \cos[ 2\pi f(t-t') ],
\label{nanb}
\end{align}
where the angle bracket denotes the ensemble average, $\delta n^i_{a}(t) = \delta n^i(t, {\boldsymbol n}_{a})$, $f_l = 1 / T$, and $f_u = 1 / 2\Delta t$. Here $\Delta t$ is the cadence of the observation, $T$ is the total observational period, and $\boldsymbol n_a$ is the sky location of the light source $a$. The power spectrum $S(f)$ and the overlap reduction function $\Gamma_{ab}^{ij}$ are defined in the frequency space as
\begin{align}
\big \langle 
{\delta n}_a^i(f) 
{\delta n}^{j*}_b(f') 
\big \rangle
= \frac{1}{2} \delta(f-f') S(f) \Gamma_{ab}^{ij},
\label{corr}
\end{align}
where $\delta n^i_a (f) = \int dt \, e^{2\pi i f t} \delta n^i_a(t)$. The spectrum is defined as a one-sided spectrum. 

The angular deflection is sourced by the underlying metric fluctuations, which are sourced by the ULDM density fluctuations. To compute the signal power spectrum, we first parameterize the metric perturbations as\footnote{As the energy-momentum tensor is quadratic in the ULDM field, vector and tensor metric fluctuations will be also generated. Since the computation of the angular deflection involves various integrals along the worldline of photon, observer, and source, it is unclear if other metric perturbations can be safely neglected. In the Appendix~\ref{app:ad_computation}, we show that the vector and tensor perturbations are indeed suppressed by the additional power of velocity in the final expression both for the fast and slow modes.}
\begin{equation}
ds^2=
- \big[1+2\Phi(t,\vec{x})\big]dt^2
+\big[1-2\Psi(t,\vec{x})\big]dx^2\,.
\end{equation}
Using \eqref{deflection_general} -- \eqref{deflection_aberration} as well as the field equations, we find approximate expressions for the angular deflection sourced by the fast and slow modes of density fluctuations as
\begin{align}
[ \widetilde{\delta n}_a (k) ]^i_F 
&= - P^{ij}(n) U_a \frac{k^j}{\omega} \widetilde{\Psi}_F(k)
\nonumber\\
&= + P^{ij}(n) U_a \frac{k^j}{\omega} \frac{4\pi G}{k^2} \widetilde{\delta \rho}_F(k)
\label{fast_dn}
\\
[ \widetilde{\delta n}_a (k) ]^i_S 
&= + P^{ij}(n) U_a \frac{k^j}{\omega} \widetilde\Phi_S (k)
\nonumber\\
&= - P^{ij}(n) U_a \frac{k^j}{\omega} \frac{4\pi G}{k^2} \widetilde{\delta \rho}_S(k)
\label{slow_dn}
\end{align}
where $U_a  = 1 - \exp[i  \omega d_a ( 1 + \boldsymbol n \cdot \boldsymbol k / \omega)]$ with $d_a$ being the distance between the observer and the source $a$. The quantities with tilde represent their Fourier component, defined as $A(x) = \int [d^4k/(2\pi)^4] e^{ik\cdot x} \widetilde A(k)$. In the second expression of each equation, we have used the Poisson equation $\nabla^2 \Psi = 4\pi G \delta\rho$ and $\widetilde\Phi_S(k) \approx \widetilde \Psi_S (k)$ for the slow modes of ULDM fluctuations~\cite{Kim:2023pkx, Kim:2023kyy}. A detailed derivation is provided in the Appendix~\ref{app:ad_computation}. 

From \eqref{corr} -- \eqref{slow_dn}, the spectrum and the overlap reduction function from ULDM fluctuations are given by
\begin{align}
&
[S(f) \Gamma_{ab}^{ij}]_{\rm DM}
\nonumber\\
= &
\frac{8G^2}{f^2}
\int \frac{d^3k}{(2\pi)^3} 
P^{ik}(\boldsymbol n_a)
P^{jl}(\boldsymbol n_b)
U_a U_b^*  \frac{ k^k k^l }{ k^4 } {\cal P}_{\delta \rho}(f, \boldsymbol k)
\nonumber\\
= &
\frac{8G^2}{3f^2}
\int \frac{dk}{2\pi^2}
P^{ik}(\boldsymbol n_a)
P^{jl}(\boldsymbol n_b)
(1+\delta_{ab}) {\cal P}_{\delta \rho}(f,k), 
\label{SG}
\end{align}
where ${\cal P}_{\delta \rho}(k)$ is the power spectrum of the density fluctuation, defined as
\begin{align}
\langle \widetilde{\delta\rho}(k) \widetilde{\delta\rho}^*(k') \rangle
= (2\pi)^4 \delta^{(4)}(k-k') {\cal P}_{\delta \rho}(k). 
\end{align}
In \eqref{SG}, we have assumed that the density power spectrum does not depend on the direction of $\boldsymbol k$ and $2\pi f d_a \gg 1$. We can then conveniently separate the power spectrum and the overlap reduction function as\footnote{We consider a few years of the observational period while assuming the typical distance to light sources $d_a \gg 1 \, {\rm pc}$. The long-distance limit $2\pi f d_a \gg1$ is justified. In addition, the $\delta_{ab}$ in the overlap reduction function is due to the perturbation at the location of the source. Since it does not give rise to a correlation of angular deflection, we ignore it for the following discussion.} 
\begin{align}
S(f)
&= 
\frac{4G^2}{\pi^2 f^2}
\int dk \, {\cal P}_{\delta \rho}(f,k) , 
\\
\Gamma_{ab}^{ij}
&= \frac{1}{3}(1+\delta_{ab}) 
P^{ik}(\boldsymbol n_a)
P^{jk}(\boldsymbol n_b) . 
\label{orf}
\end{align}
The overlap reduction function $\Gamma_{ab}^{ij}$ is common for both the fast and slow modes. 

In the following subsections, we summarize the signal spectrum from the fast and slow modes of ULDM fluctuations. For simplicity, we assume the normal dark matter velocity distribution,
$$
f(v)
= \frac{\bar\rho/m}{(2\pi \sigma_v^2)^{3/2}} 
\exp\left[ - \frac{v^2}{2\sigma_v^2} \right] ,
$$
with a mean density $\bar\rho$ and the velocity dispersion $\sigma_v$. We then use the power spectrum for the density fluctuation ${\cal P}_{\delta\rho}(k)$ presented in Appendix~\ref{app:ULDM_computation}.

\subsection{Fast Mode}
Using the power spectrum of the fast mode ULDM density fluctuation \eqref{fast_density_fluctuation}, we find the signal spectrum of the fast mode as
\begin{align}
S(f) 
&= 
\frac{\pi}{2 f^2} \frac{(G\rho)^2}{m^3}
\frac{\bar v^6}{\sigma_v^6}
\exp\Big( - \frac{\bar v^2}{\sigma_v^2} \Big) , 
\label{fast_spectrum}
\end{align}
where $\bar v^2(f) = 2\pi f/m - 2$, and $\sigma_v = 160\,{\rm km/sec}$ is the velocity dispersion of dark matter. The spectrum strongly peaks around $f_m =m/\pi$ with a width $\Delta f = m \sigma_v^2/(2\pi)$. Since the width is much smaller than the peak frequency, one may replace $1/f^2 = 1/f_m^2$ without significant quantitative changes. 

When $T < 1 / \Delta f$, the lineshape of the signal cannot be resolved. The total power is concentrated at a single frequency bin centered at $f = f_m$. In this case, the power spectrum can be approximated as $S(f) \to \delta(f-f_m) \int_{f_m}^\infty df \, S(f)$, i.e.
\begin{align}
S(f) \approx \frac{3\pi}{2} \frac{(G\rho\sigma_v)^2}{m^4} \delta(f-f_m). 
\label{sigma_f_approx}
\end{align}
For benchmarks we consider below, this limit $T < 1 / \Delta f$ is always satisfied for the fast mode. We use the above approximate expression for the spectrum to estimate the sensitivities in the following section.  

\subsection{Slow Mode}
Using the power spectrum of the slow mode ULDM density fluctuation \eqref{slow_density_fluctuation}, we find the signal spectrum of the slow mode as
\begin{align}
S(f) 
= 
\frac{8}{f^2} \frac{(G \rho)^2}{m^3 \sigma_v^4} 
K_0\left( \left| \frac{2\pi f}{m \sigma_v^2}\right| \right)  , 
\label{slow_mode}
\end{align}
where $K_n(x)$ is the modified Bessel function of the second kind. The spectrum is exponentially suppressed for $f > m \sigma_v^2 / 2\pi$, while it logarithmically increases $S(f) \propto \log(1/f)$ for $f < m \sigma_v^2/2\pi$. The exponential suppression at $f > m \sigma_v^2/2\pi$ is due to the underlying dark matter velocity distribution, while the logarithmic dependence is due to the long-range nature of the gravitational interaction.

\section{Sensitivity Estimation}\label{sec:snr}
From measurements of each light source, the angular position of the background light source is determined and the astrometric solution is found. The difference between the observed position and the position predicted by the astrometric solution is the residual fluctuations, parameterized as
\begin{align}
\delta n^i_a(t) 
= s_a^i(t) 
= r_a^i(t) + h_a^i(t) , 
\end{align}
where $r^i_a(t)$ is noise, and $h^i_a(t) = h^i(t, \boldsymbol n_a)$ is the signal, either due to gravitational waves or ultralight dark matter or both. In the following sections, we estimate the sensitivity of the current and future astrometric observations for ultralight dark matter using the cross-correlation test statistic. 

\subsection{Correlation}
The angular deflection forms a vector field in the sky. At a given point in the sky $\hat n$, the angular deflection can be described by a two-component vector. In particular, one may expand it as $\delta n = \delta n_\theta \hat \theta + \delta n_\phi \hat\phi$, where $(\hat n, \hat\theta,\hat \phi)$ forms an orthonormal basis in the spherical coordinate system. For ease of notation, we aggregate the indices as
$$s_a^i (t) \to s_A(t) , $$ 
where the collective index $A$ denotes $(ai)$. Note that the index $i$ spans a tangent plane at the sky position $\hat n_a$. With this new notation, the noise and signal correlators are given by
\begin{align}
\langle \tilde r_A(f) \tilde r_{B}^*(f') \rangle
&= \frac{1}{2} \delta(f-f') \delta_{AB} S_A(f), 
\label{noise_corr}
\\
\langle \tilde h_A(f) \tilde h^*_B(f') \rangle
&= \frac{1}{2} \delta(f-f') S_{AB}(f) .
\label{signal_corr}
\end{align}
We assume that the noise is uncorrelated over different background light sources and each direction in the tangent plane. For the ULDM signal, the $S_{AB}(f)$ is related to the signal power spectrum and the overlap reduction function as
\begin{align}
S_{AB}(f) = [S(f) \Gamma_{AB}]_{\rm DM}.
\end{align}
The overlap reduction function should be understood as $\Gamma_{AB}  = \Gamma^{ij}_{ab}$.

\subsection{Signal-to-Noise Ratio}
To estimate the sensitivity, we use the optimal cross-correlation statistic~\cite{Allen:1997ad, Anholm:2008wy}
\begin{align}
\hat Y = 
2\, {\rm Re}\,
\sum_{A \neq B} 
\int_{f_l}^{f_u} df \, \tilde s_A(f) \tilde s_B^*(f) \frac{S_{AB}(f)}{S_A(f) S_B(f)} ,
\label{ts}
\end{align}
where $S_{AB}(f) / S_{A}(f) S_B(f)$ in the integrand is an optimal filter in the weak signal limit. Since we are interested in ULDM signals, we choose $S_{AB} = [S(f) \Gamma_{AB}]_{\rm DM}$. 
The dark matter signal is then characterized by the mean value of the statistic in the presence of the signal $\mu_{\hat{Y}} = \langle \hat{Y} \rangle_{r+h}$. At the same time, the noise is characterized by the variance of the statistic in the absence of the signal, $\sigma^2_{\hat{Y}} = \langle \hat{Y}^2 \rangle_{r} - \langle \hat{Y} \rangle^2_{r}$. The detection statistic can be constructed as
\begin{align}
\hat\rho = \frac{\hat{Y}}{\sigma_{\hat Y}}  ,
\label{dts}
\end{align}
and the signal-to-noise ratio can then be estimated as
\begin{align}
 {\rm SNR}_{\rm DM}^2
=\langle \hat\rho \rangle^2
= 
\sum_{A \neq B}
T \int_{f_l}^{f_u} df \, \frac{S_{AB}^2(f)}{S_A(f) S_B(f)}
\label{SNR}
\end{align}
where the upper and lower limits are fixed by the inverse of the cadence $f_u = 1/ 2\Delta t$ and the total observational time scale $f_l=1/T$, respectively. 

An approximate expression for the signal-to-noise ratio can be obtained by substituting the explicit expressions for the ULDM spectrum and overlap reduction function. Assuming that the noise is white, we find 
\begin{align}
{\rm SNR}_{\rm DM}^2
&= 
\frac{ T }{ (\Delta t)^2 }
\sum_{A \neq B} \frac{ \Gamma_{AB}^2 }{ \sigma_A^2 \sigma_B^2 } 
\int_{f_l}^{f_u} df \, S_{\rm DM}^2(f)
\nonumber\\
& \approx
\frac{4T}{27(\Delta t)^2}
\frac{N_\star^2}{\sigma_r^4} 
\int_{f_l}^{f_u} df \, S_{\rm DM}^2(f) , 
\label{snr}
\end{align}
where we use $S_A = \Delta t \sigma_A^2$ for the auto-correlation and approximate $\sigma_A^2 = \sigma_B^2 = \sigma_r^2$ for all stellar populations in the second line. In addition, we approximate the squared overlap reduction function to its solid angle-averaged value, $\Gamma_{AB}^2 \to \langle \Gamma_{AB}^2 \rangle = 4/27$. 

With the fast and slow mode power spectra \eqref{sigma_f_approx}--\eqref{slow_mode}, the above signal-to-noise ratio can be computed. For the slow mode, one must perform the full frequency integral, while, for the fast mode, the integral is performed with an approximate expression for the spectrum \eqref{sigma_f_approx} as the coherence time of the fast mode signal is much longer than the typical observational time scale.

\subsection{Results}

\begin{table}[t]
\centering
\bgroup
\renewcommand{\arraystretch}{1.5}
\setlength\tabcolsep{6.pt}
\begin{tabular}{ccccc}
\toprule
& $T$ 
& $\Delta t$ 
& $\sigma_r$ 
& $N_\star$
\\ \hline 
BM1 
& $10$-yr
& $4$ weeks
& $100\,\mu$as
& $10^8$
\\
BM2 
& $10$-yr
& $15$ mins
& $100\,\mu$as
& $10^7$
\\
\bottomrule
\end{tabular}
\egroup
\caption{Two benchmarks used for the sensitivity projections. The $T$ is the observational period, $\Delta t$ is the observational cadence, $\sigma_r$ is the single-exposure position uncertainty, and $N_\star$ is the number of sources. The first benchmark is motivated by the current Gaia mission, while the second benchmark is motivated by the Galactic Bulge Time Domain Survey of the Roman Space Telescope.}
\label{tab:bms}
\end{table}

\begin{figure*}[t]
\centering
\includegraphics[width=0.75\textwidth]{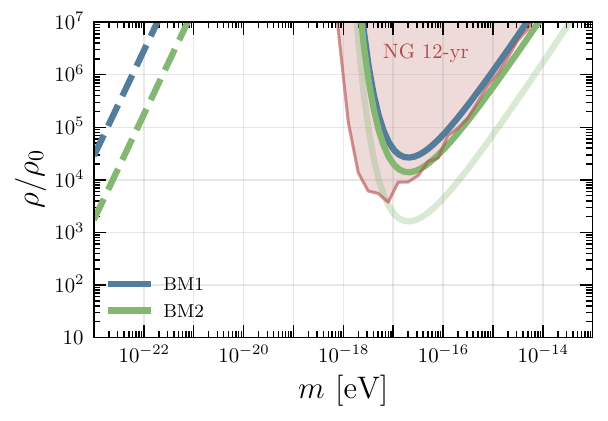}
\caption{The projected sensitivities of two benchmarks in Table~\ref{tab:bms}. The dashed lines are based on the coherent signals from the fast mode fluctuations at $\omega = 2m$, while the solid lines are based on the stochastic signals from the slow mode fluctuations at $\omega \lesssim m \sigma_v^2$. The light green line is the projection obtained directly from \eqref{snr}, while the dark green line is obtained assuming that the survey observes only $\Delta \Omega \sim (1.5\,{\rm deg})^2$ of the entire sky and that no correlated noise induced by the stochastic gravitational wave background can be subtracted from the data. See Section~\ref{sec:gw_noise} for the detailed discussion. The result from the analysis of the NANOGrav 12.5-year dataset is reproduced from Ref.~\cite{Kim:2023kyy}. }
\label{fig:result}
\end{figure*}

Using the signal-to-noise ratio obtained in the previous section, we project the sensitivities of astrometric observations for ultralight dark matter search. For this purpose, we choose two benchmark specifications summarized in Table~\ref{tab:bms}. These two benchmarks are chosen to show approximate sensitivities of the current and future observations such as Gaia and the Roman Space Telescope. The first benchmark is motivated by the Gaia mission. The Gaia mission is already extended to the end of 2025, and the final data release is expected to contain 10 years of observation data. During the mission lifetime, each light source is visited $\sim 140$ times on average~\cite{2021A&A...652A..76H}, which translates into about a month of observation cadence. While the position error depends on the magnitude of the source, we choose $\sigma =100\,{\mu \rm as}$. This choice will be further explained and justified in Section~\ref{label:gaia_rel}. 

The second benchmark is motivated by the Galactic Bulge Time Domain Survey of the Roman Space Telescope. The survey intends to observe $N_\star \sim 10^8$ stars down to the magnitude $H_{AB}=21.6$ over 6 seasons of 72 days during 5 years of the nominal mission lifetime~\cite{2019ApJS..241....3P, Roman_Surveys}. The observational cadence is $15$-minute. Given the currently available specification of the mission, we choose $\Delta t = 15\,{\rm min}$, $\sigma = 100\,{\mu \rm as}$, and $N_\star = 10^7$.

We present the projected sensitivities for each benchmark in Figure~\ref{fig:result}. The dashed lines are based on the coherent signal at $\omega =2m$ (fast mode), while the solid lines are based on the stochastic signal at $\omega \lesssim m \sigma_v^2$ (slow mode). The result is normalized with a local dark matter density $\rho_0 \simeq 0.4\,{\rm GeV/cm^3}$. We choose ${\rm SNR}=1$ for the projections.  

We show two projections for the ULDM stochastic fluctuation search with the second benchmark. The light green line is derived from the signal-to-noise ratio derived in the previous section \eqref{snr}, which assumes a full sky survey. The solid green line is a projection, considering that one surveys only a small fraction of the sky, e.g. $\Delta \Omega = (1.5\,{\rm deg})^2$, that the stochastic gravitational wave background inferred from recent pulsar timing array observations constitutes additional correlated noise source, and that none of such correlated noise can be subtracted. A detailed discussion on this point is presented in the section~\ref{sec:gw_noise}.

\section{Discussion}\label{sec:discussion}

\subsection{Acceleration of the Solar System}
The angular deflection due to ULDM fluctuation may be expanded in terms of vector spherical harmonics. As shown in Appendix~\ref{app:vsh}, the ULDM-induced signal gives rise to a dipole correlation pattern ($\ell=1$) over the sky. It is therefore worthwhile to explore if there are any other systematic effects at the dipole. 

The acceleration of the solar system towards the galactic center indeed provides a dipole pattern of the angular deflection via the light aberration. As the solar system is accelerated due to the galactic potential, it provides a drift of angular deflection at the level of $\dot{\delta n} = P^{ij}(n) a^i \sim 5 \,{\rm \mu as}/{\rm yr}$. This effect has been measured via Very Long Baseline Interferometry (VLBI) observations of quasi-stellar objects (QSOs)~\cite{Gwinn:1996gv, Titov:2010zn, 2020A&A...644A.159C} and more recently via precision astrometric observations of QSO-like objects in Gaia EDR3~\cite{2021A&A...649A...9G}. The effect from the solar system acceleration can be therefore modeled and subtracted when searching for ultralight dark matter. In addition, there could be acceleration from the Galactic bar and spirals, acceleration toward the Galactic plane, and acceleration due to nearby stars, molecular clouds, and so on. These additional accelerations would provide similar dipole-like angular deflection on the sky, and therefore, constitute unwanted systematic effects already for the measurement of the acceleration of the solar system. In Ref.~\cite{2021A&A...649A...9G}, these systematic uncertainties are estimated and shown to be at most at the level of ${\cal O}(0.1)\,{\rm \mu as/yr}$. In addition, given that these uncertainties provide a drift of angular deflection that is relevant over a time scale longer than a few years, we expect that it would not directly affect the sensitivity of ultralight dark matter searches as we focus on time-dependent signatures of ULDM signals at time scales smaller than a few years.

\subsection{Relative Astrometry}
The second benchmark in Table~\ref{tab:bms} is motivated by the Galactic Bulge Time-Domain Survey of the Roman Space Telescope. This survey will provide a relative astrometry of about a billion stars near the center of our Galaxy. In particular, it will observe $2\,{\rm deg}^2$ at the Galactic bulge at a 15-minute cadence over 6 seasons each lasting 72 days. 

The ultralight dark matter signal is spread over a large angular scale. Especially, the vector spherical harmonics expansion shows that the signal is dominantly dipole. It is therefore conceivable that the ULDM signal is suppressed, especially when the only available measurement is the relative astrometric positions of stars with respect to a certain reference star within a narrow range of the sky. The same applies to the stochastic gravitational wave since the most power of the stochastic gravitational wave background resides in the quadrupole. The potential suppression of SGWB signals in relative astrometry observation was discussed in Ref.~\cite{Wang:2020pmf}, and several mitigation strategies were discussed in Refs.~\cite{Wang:2022sxn, Pardo:2023cag}. 

Relative astrometry observations could be converted into absolute ones. Two possibilities were explored in previous literature~\cite{2019JATIS...5d4005W, WFIRST-STScI-TR1702}. One way, called {\it a prior} method, relies on the available information before the observation. The telescope uses at least four guiding stars for the telescope pointing. With prior information on the celestial coordinates of the guiding stars, the relative position of galactic bulge stars can be converted into the absolute position. It has been shown that, within the telescope's field-of-view (FoV), there exist about a thousand guiding stars on average over the entire sky based on Two Micron All Sky Survey (2MASS) catalog~\cite{WFIRST-STScI-TR1702}, which enables the conversion of relative astrometry to the absolute one. 

Another method is referred to as {\it a posterior} method. In this method, the relative position of stars is converted into their absolute position after the observation has been made by analyzing all available faint stars within FoV whose position is accurately known in external catalogs such as Gaia. It has been demonstrated, based on the observation of Gaia during its first year, that the available number of Gaia stars near the galactic center will be more than a few thousand~\cite{WFIRST-STScI-TR1702}. This conversion of the relative astrometry to the absolute one, in conjunction with the external catalog, could potentially allow us to circumvent the limitations of partial sky coverage of the Galactic Bulge Time-Domain Survey.

\subsection{Populations with Different Position Errors}\label{label:gaia_rel}
In the signal-to-noise estimation, we assume that all background light source has the same single-exposure position uncertainty $\sigma_r$. This is an unrealistic assumption as the error depends on the magnitude of the light source. The purpose of this section is to examine if the benchmark, especially the first one, is realistic given the current available information on the Gaia observation.

The current data release of Gaia provides astrometric solutions for $N_\star \simeq 1.8\times 10^9$ background light sources~\cite{2023A&A...674A...1G}. The typical error in the astrometric observation depends on the magnitude of the stars. The distribution of the magnitude of stars is provided in Figure~1 of Ref.~\cite{2023A&A...674A...1G}, while the predicted parallax errors for stars in each magnitude are summarized in~\cite{gaia_science_performance}. For instance, with Gaia DR4, the expected parallax error at $G=18$ is $\sigma_{\varpi} = 100\mu{\rm as}$. The parallax error is then converted into the sky-averaged position error using $\sigma_0 = 0.75\sigma_{\varpi}$. To convert this number to the single-exposure position error, we inflate the number as $\sigma_r = \sqrt{N_{\rm obs}} \sigma_0$, where $N_{\rm obs}$ is the number of observations for a given star. As DR4 will be based on 66 months of observations, we use $N_{\rm obs} =66$ assuming an average observational cadence of a month. With this conversion, we find, for instance, $\sigma_r(G=13) = 59\,{\mu \rm as}$, $\sigma_r(G=15) = 135\,{\mu \rm as}$, $\sigma_r(G=17) = 361\,{\mu \rm as}$, $\sigma_r(G=19) = 1.3\,{\rm mas}$, and $\sigma_r(G=21) = 6.6\,{\rm mas}$ as a single-exposure position error at each magnitude. 

\begin{figure}
\includegraphics[width=0.45\textwidth]{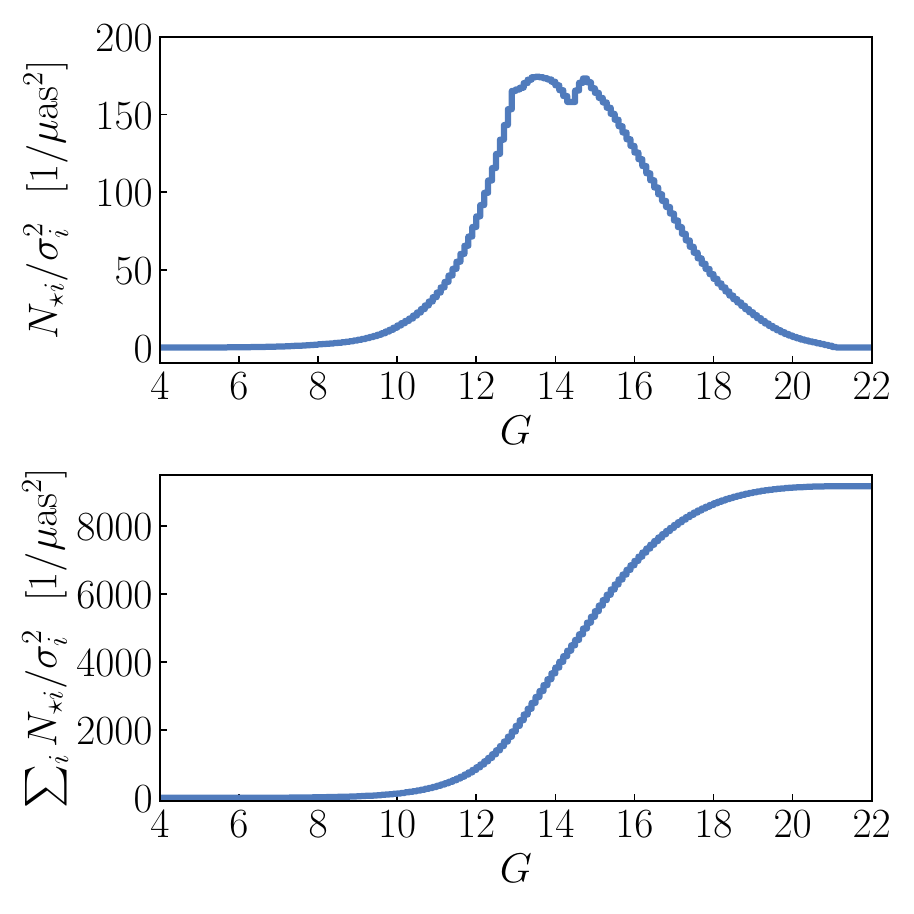}
\caption{(Top) the distribution of $N_{\star i}/\sigma_i^2$ as a function of the magnitude. The number of star $N_{\star i}$ is given in $\Delta G = 0.1$ mag bin. (Bottom) the cumulative sum of $N_{\star i}/\sigma_i^2$. }
\label{fig:sub}
\end{figure}

As realistic data contains populations of stars with different magnitudes and position errors, the signal-to-noise estimation changes accordingly. In the signal-to-noise ratio \eqref{snr}, 
\begin{align}
\frac{N_\star}{\sigma^2}
\to \sum_i 
\Big( \frac{N_{\star}}{\sigma^2} \Big)_i, 
\end{align}
where the sum is performed over each magnitude bin. With the $G$-magnitude distribution and estimated single-exposure errors, we show in Figure~\ref{fig:sub} the distribution of $N_{\star i}/\sigma_{i}^2$ and its cumulative sum. The sum leads to $\sum_i N_{\star i}/\sigma_i^2 \sim 9\times 10^3 \, \mu{\rm as}^{-2}$. In the first benchmark, we choose $N_\star =10^8$ and $\sigma_r = 100\,{\rm \mu as}$, which leads to $N_\star / \sigma_r^2 = 10^4 {\rm \mu as}^{-2}$; this suggests that, while the chosen position error in the benchmark looks somewhat smaller than the one in the catalog, this benchmark would still provide a reasonable estimation for the sensitivity of the Gaia mission as we discount the number of available sources.

The same argument might apply to the second benchmark, which is motivated by the Galactic Bulge Time Domain Survey of Roman. The survey intends to observe $N_\star \sim 10^8$ stars down to $H_{AB} =21.6$, while the single-exposure position error at this magnitude is estimated to be $1\,{\rm mas}$~\cite{2019JATIS...5d4005W}. Similar to the first benchmark, we choose the position error somewhat smaller than the position error estimate of $H_{AB}=21.6$ stars, while discounting the number of available sources by a factor of few, hoping that this provides a reasonable estimate of the survey for the ultralight dark matter search. We acknowledge that, without knowing the magnitude distribution of stars and their corresponding position errors, it is difficult to make a quantitative statement as to whether this benchmark would provide a reasonable estimate. The result with this benchmark should be understood with this caveat.

\begin{figure}[t]
\centering
\includegraphics[width=0.45\textwidth]{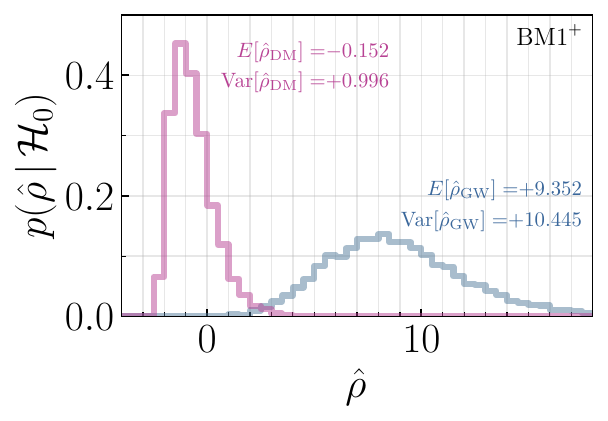}
\caption{The distribution of $\hat\rho_{\rm DM}$ and $\hat \rho_{\rm GW}$ under the hypothesis ${\cal H}_0$ with white noise and SGWB. We increase the number of light sources in BM1 to $N_\star =5 \times 10^9$. With this, one finds $\langle \hat\rho_{\rm GW} \rangle \sim 10$ analytically, which is consistent with the numerical result. The presence of SGWB does not affect the distribution of $\hat \rho_{\rm DM}$ significantly due to the non-overlap of overlap reduction functions between ULDM and SGWB. A small bias in the $\hat\rho_{\rm DM}$ can be reduced by increasing the number of pixels.}
\label{fig:os1}
\end{figure}

\subsection{Gravitational Wave Noise}\label{sec:gw_noise}
The sensitivity curves are obtained with the signal-to-noise ratio of the statistic \eqref{ts}. In doing so, we assume that the noise is uncorrelated over the different light sources. This is crucial; any correlated noise source could significantly affect the sensitivity of astrometric observations for the ultralight dark matter. 

One apparent source of the correlated noise is the stochastic gravitational wave background (SGWB). Recently pulsar timing array (PTA) collaborations have found evidence of the stochastic gravity wave background at nano-Hertz frequencies~\cite{NANOGrav:2023gor, Antoniadis:2023rey, Reardon:2023gzh, Xu:2023wog}. The inferred SGWB spectrum is consistent with an expected astrophysical source of supermassive black hole binaries. The spectrum can be described by the characteristic strain,
$$
h_c(f) = A_{\rm GWB} \left( \frac{f}{f_{\rm ref}} \right)^\alpha, 
$$
where $A_{\rm GWB}$ is the amplitude, $f_{\rm ref}$ is an arbitrary reference frequency, and $\alpha$ is the spectral index. While detailed values of $A_{\rm GWB}$ and $\alpha$ vary, they are generally consistent across different PTA analyses. 

It is therefore important to assess how additional correlated gravitational wave noise could affect the projections for the ultralight dark matter search. In particular, we ask if SGWB introduces a systematic bias in a test statistic for the ULDM search. For the investigation, we simulate angular deflections both with white noise and stochastic gravitational waves. With simulated datasets, we compute the distribution of the statistic \eqref{dts} with the ULDM-induced cross-correlation $S_{AB} = [S(f) \Gamma_{AB}]_{\rm DM}$. Without ULDM and SGWB signals, we expect $\langle \hat\rho_{\rm DM} \rangle = 0$ and ${\rm Var}(\hat\rho_{\rm DM}) =1$. The main question is if this remains the same in the presence of the injected SGWB. 

To generate a synthetic dataset, we divide the sky into a uniformly distributed equal area of $N_{\rm pix} = 192$ pixels and treat each of them as an individual light source. At each pixel, we contaminate the angular deflection with random Gaussian white noise, with an adjusted single-exposure error $ \sigma_r \sqrt{N_{\rm pix} / N_\star}$. Then, we inject SGWB signals with $A_{\rm GWB} = 6\times 10^{-15}$ and $\alpha = -0.15$. The chosen parameters for SGWB are consistent with the ones from the NANOGrav collaboration~\cite{NANOGrav:2023gor}.

The first benchmark is not sensitive enough to measure the injected SGWB. In particular, we find $\langle \hat \rho_{\rm GW} \rangle \simeq 0.2$, where the statistic $\hat\rho_{\rm GW}$ is defined the same way but with $S_{AB}(f) = [S_h(f) \Gamma_{AB}]_{\rm GW}$. As the signal-to-noise ratio is smaller than unity, we expect that it would not produce significant bias for $\hat\rho_{\rm DM}$. To demonstrate this even more clearly, we inflate the number of light sources in BM1 to $N_\star =5\times 10^9$ (denoted as BM1$^+$) and compute the distribution of $\hat\rho_{\rm DM}$ as well as $\hat\rho_{\rm GW}$. In Figure~\ref{fig:os1}, we show the distribution of both statistics; we find $\langle \hat \rho_{\rm GW} \rangle \sim 10$ as expected, while $\langle \hat \rho_{\rm DM} \rangle \simeq 0$ and ${\rm Var}(\hat\rho_{\rm DM}) =1$. This suggests that, even when the injected SGWB can be measured with BM1$^+$, it does not affect the distribution of $\hat\rho_{\rm DM}$. 

This conclusion is due to the non-overlap of the overlap reduction functions between ULDM and SGWB. When SGWB is present, the mean value of the statistic $\hat\rho_{\rm DM}$ is
\begin{align}
\langle \hat \rho_{\rm DM} \rangle
\propto \sum_{A\neq B} \Gamma_{AB}^{\rm GW} \Gamma_{BA}^{\rm DM},
\label{overlap_overlap}
\end{align}
where $\Gamma_{AB}^{\rm GW}$ and $\Gamma_{AB}^{\rm DM}$ are the overlap reduction functions for SGWB and ULDM, respectively. In the vector spherical harmonics expansion, $\Gamma_{AB}^{\rm GW}$ has non-vanishing components only for $\ell \geq2$~\cite{Book:2010pf}, while $\Gamma_{AB}^{\rm DM}$ is non-vanishing for $\ell =1$ (App.~\ref{app:vsh}). In the limit where the sum becomes the continuous integral over the sphere $S^2$, the sum vanishes due to the orthogonality of the vector spherical harmonics. With the finite number of pixels, a small bias could be introduced as is already present in the distribution of $\hat\rho_{\rm DM}$ in Fig.~\ref{fig:os1}, but the bias can be further reduced by increasing the number of pixels in the analysis. 

\begin{figure}
\centering
\includegraphics[width=0.45\textwidth]{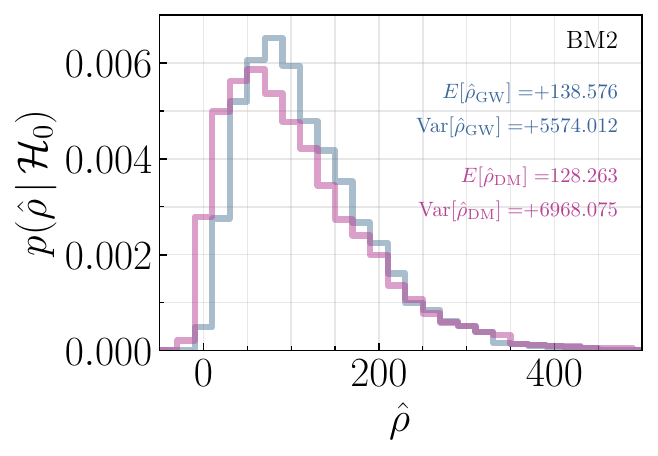}
\caption{Same as Figure~\ref{fig:os1} but with BM2. The background light sources are assumed to be centered around the galactic bulge with $\Delta \Omega =(1.5\,{\rm deg})^2$. The injected SGWB introduces a significant systematic bias in $\hat\rho_{\rm DM}$.}
\label{fig:os2}
\end{figure}

This discussion raises a more serious concern for the projection based on the second benchmark. If we take the second benchmark to mirror the Galactic Bulge Time-Domain Survey of Roman, the sum \eqref{overlap_overlap} only covers $\sim 2\,{\rm deg}^2$ of the sky near the galactic bulge. As the survey region is too narrow to disentangle the ULDM dipole signals from SGWB quadrupole noises, the sum will introduce a significant systematic bias as is shown in Fig.~\ref{fig:os2}.

If none of the SGWB-induced correlated noise can be subtracted, the signal-to-noise ratio decreases accordingly. The signal-to-noise ratio in the presence of correlated SGWB noise is given by (App.~\ref{app:os})
\begin{align}
\widetilde {\rm SNR}_{\rm DM}
= \left[ T \int_{f_l}^{f_u} df \, \Tr\big(\tilde \Sigma^{-1} \, S^{\rm DM} \, \tilde\Sigma^{-1} \, S^{\rm DM} \big) \right]^{\frac12}, 
\end{align}
where $\tilde \Sigma_{AB}(f) = S_{A}(f) \delta_{AB} + S_{AB}^{\rm GW}(f)$ is the noise covariance matrix in the presence of SGWB. Assuming that the white noise covariance matrix is proportional to the diagonal matrix $S_{A}\delta_{AB} = S_N \delta_{AB}$, and ${\rm SNR}_{\rm GW} \gg 1$, the signal-to-noise ratio behaves parametrically 
\begin{align}
\widetilde{\rm SNR}_{\rm DM} 
\sim \frac{{\rm SNR}_{\rm DM}}{{\rm SNR}_{\rm GW}} ,
\end{align}
where ${\rm SNR}_{\rm DM}$ is the signal-to-noise ratio without the correlated gravity wave noise. As the signal-to-noise ratio for the SGWB with the second benchmark is about ${\rm SNR}_{\rm GW} \sim 10^2$, this results in an order of magnitude suppressed sensitivity for $\rho/\rho_0$. We numerically compute the ratio $\widetilde {\rm SNR}_{\rm DM} / {\rm SNR}_{\rm DM}$ with the injected SGWB, and show the resulting projection on $\rho/\rho_0$ as a solid green line in Fig.~\ref{fig:result}. 

The correlated noise might be subtracted to some degree. For SGWB searches in a global network of ground-based gravity wave detectors, it has been recognized that correlated magnetic noise from Schumann resonances could contaminate observations, thereby challenging the detection of SGWB~\cite{Christensen:1992wi, Allen:1997ad, Thrane:2013npa}. One possible way to mitigate correlated noise is to use local magnetometers as supplementary witness channels and subtract their measurements from the gravitational wave strain data channels coherently~\cite{Thrane:2014yza, Coughlin:2016vor, Himemoto:2017gnw, Coughlin:2018tjc, Himemoto:2019iwd, Meyers:2020qrb}. This so-called Wiener filtering scheme, as already demonstrated in Refs.~\cite{Coughlin:2016vor, Coughlin:2018tjc}, could effectively mitigate correlated noise in spatially separated magnetometers to the level of uncorrelated noise. 
As we could use pulsar timing arrays as an additional witness channel, the correlated SGWB noise in astrometric observation, especially the ones with $\Delta \Omega \ll1$, might be, at least partially, subtracted. The resulting projection would be somewhere between the light and dark green lines, depending on the level of subtraction. To determine the degree of possible subtraction, a further dedicated investigation is required. We leave it for future study.

\subsection{Local Dark Matter Density}
The result shown in Figure~\ref{fig:result} is normalized with the local dark matter density $\rho_0 = 0.4\,{\rm GeV/cm^3}$. This local dark matter density is inferred from the investigation of tracers over the spatial volume of $V \sim {\cal O}({\rm kpc}^3)$ near the solar system (see e.g. Refs.~\cite{Read:2014qva, deSalas:2020hbh} for reviews). On the other hand, the current analysis of astrometric surveys of ULDM would be sensitive to the very local dark matter density, a few hundred AU within the solar system, as the main effect arises from the interaction between the observer and the ultralight dark matter, and as the solar system only sweeps through the dark matter halo about a few hundred AU scale over the time scale of ten years. At this small spatial scale,  direct gravitational measurements of dark matter are absent, with only constraints currently in place. Several existing constraints include the one from the analysis of solar system ephemerides $\rho/\rho_0 < 2\times 10^4$~\cite{Pitjev:2013sfa}, lunar laser ranging and geodetic satellites LAGEOS $\rho/\rho_0 < 10^{11}$~\cite{Adler:2008rq}, and the motion of asteroids in the solar system $\rho/\rho_0 < 6\times 10^6$~\cite{Tsai:2022jnv}. Particularly, for ULDM, a recent analysis of pulsar timing array has placed a strong constraint $\rho/\rho_0 < 4\times 10^3$ at $m \simeq 10^{-17}\eV$~\cite{Kim:2023kyy}. In this regard, we expect that the astrometric survey will provide the strongest constraint around $m \sim 10^{-17}\eV$ along with the pulsar timing array.

\section{Conclusion}\label{sec:conclusion}
In this work, we have investigated the sensitivities of astrometric observations for the ultralight dark matter search via gravitational interaction. By computing the angular deflection in the presence of a generic metric fluctuation, we characterize the effects of ultralight dark matter in astrometry observations using its signal spectrum and the overlap reduction function. With two benchmarks that mimic the designs of current and future astrometric observation by the Gaia and Roman Space Telescope, we have demonstrated that $\rho/ \rho_0 \simeq {\cal O}(10^{3\textrm{--}4})$ could be probed at the mass range $10^{-18}\eV \lesssim m \lesssim 10^{-16}\,{\rm eV}$. This sensitivity is comparable to the sensitivity of the current pulsar timing array. 

We have only considered the gravitational interaction between ultralight dark matter and the standard model particles. An interesting future direction would be to investigate the implications of non-gravitational interaction between ultralight dark matter and SM particles. Depending on the nature of microscopical interaction, it might lead to absorption or emission of ultralight dark matter or an elastic scattering between the density fluctuations and SM particles. This possibility will be investigated in future work.

\begin{acknowledgments}
We thank Andrea Mitridate and Xiao Xue for useful conversations and comments on the manuscript. While this work is being finalized, we learned of the related work~\cite{DrorVerner}, which also considers an astrometric search for ultralight dark matter. We thank Jeff Dror and Sarunas Verner for coordinating the submission. This work was supported by the Deutsche Forschungsgemeinschaft under Germany’s Excellence Strategy - EXC 2121 Quantum Universe - 390833306.
\end{acknowledgments}

\bigskip

\appendix
\section{ULDM Fluctuations}\label{app:ULDM_computation}
\subsection{Density}
We summarize the properties of ultralight dark matter density fluctuations, following~\cite{Kim:2023pkx, Kim:2023pvt, Kim:2023kyy}. We consider a free massive scalar field whose action is given as
\begin{align}
S = \int d^4x \, 
\left[ 
- \frac{1}{2} \eta^{\mu\nu} \partial_\mu \phi \partial_\nu \phi 
- \frac{1}{2}m^2\phi^2 
\right].
\end{align}
Note that we adopt the mostly positive metric signature, $\eta^{\mu\nu}= ( - + + + )$. Each element of the energy-momentum tensor is considered as the first order in the perturbation theory, and therefore, we use the flat spacetime metric for the following computation. The scalar field is expanded in non-relativistic limit as\footnote{More precisely, $1/\sqrt{2mV} \to 1/\sqrt{2\omega_i V}$. This does not affect the results that are leading order in the non-relativistic expansion.}
\begin{align}
\phi = \frac{1}{\sqrt{2mV}} \sum_i 
[ a_i e^{i k_i \cdot x} + a_i^* e^{- i k_i \cdot x} ],
\end{align}
where $i$ denotes any conserved quantum numbers of the system. In our case, it is simply a three-momentum of dark matter particles. Here $(a_i, a_i^*)$ is a set of complex random numbers with the probability density function $p(a_i)$.

We assume that the field $\phi$ is a Gaussian random field. The probability distribution of the complex random number $a_i = r_i e^{i\theta_i}$ is then given by~\cite{Derevianko:2016vpm, Foster:2017hbq, Centers:2019dyn, Kim:2021yyo} 
\begin{align}
dP 
= d^2 a_i \, p(a_i)
= d^2 a_i \, \left[
\frac{1}{\pi f_i} 
\exp\left(
- \frac{|a_i|^2}{f_i}
\right)
\right], 
\end{align}
where $f_i$ is the phase space distribution of the mode $i$ and the probability measure is defined as $d^2\alpha_i = dr_i r_i d\theta_i$. The phase is uniformly distributed, while the amplitude follows the Rayleigh distribution. The ensemble average of any combination of scalar field operators can be computed as
\begin{align}
\Big\langle O[\phi] \Big\rangle
=\int \Big[ \prod_{i} d^2 a_i \, p(a_i) \Big] O[\phi]. 
\end{align}

The density of the scalar field is $\rho = [\dot\phi^2 + (\nabla \phi)^2 + m^2\phi^2 ]/2$. It can be decomposed into the fast and slow mode,
$$
\rho (x) 
= \rho_F(x) 
+ \rho_S(x) , 
$$
where each of them is given by
\begin{align}
\rho_F(x)
&=
\frac{m}{2V} \sum_{1,2} 
R_{12}^{F}
\Big[ 
a_1 a_2 e^{i( k_1 + k_2) \cdot x}
+{\rm h.c.} \Big] , 
\\
\rho_S(x)
&= 
\frac{m}{2V} \sum_{1,2} 
R_{12}^{S} 
\Big[ 
a_1 a^*_2 e^{i( k_1 - k_2 ) \cdot x}
+{\rm h.c.} \Big] . 
\end{align}
The fast mode oscillates at $\omega = \pm( \omega_{1} + \omega_2) \simeq \pm 2m$, while the slow mode oscillates at $\omega = \pm  (\omega_1 - \omega_2) \sim \pm m \sigma_v^2$. The coefficient $R_{12}^{F, S}$ is defined as
\begin{align}
R_{12}^{F}
&= 
\frac{1}{2}
- \frac{ \omega_1 \omega_2 + \boldsymbol k_1 \cdot \boldsymbol k_2 }{2m^2} 
\simeq
- \frac{( \boldsymbol v_1 + \boldsymbol v_2)^2}{4},
\\
R_{12}^{S}
&= 
\frac{1}{2} + \frac{\omega_1 \omega_2 + \boldsymbol k_1 \cdot \boldsymbol k_2}{2m^2}
\simeq  1.
\end{align}
The second expression in each line is obtained in the non-relativistic limit. The fast mode is suppressed by the velocity square compared to the slow mode.

The mean value of the density of each mode is then given by
\begin{align}
\langle \rho_F(x) \rangle &= 0 , 
\\
\langle \rho_S(x) \rangle &= m \int d^3 v \, f(\boldsymbol v) 
\equiv \bar\rho , 
\end{align}
where we take the continuum limit in the second line, $\sum_i (f_i/V) \to \int d^3 v \, f(\boldsymbol v)$, with the continuum velocity distribution of dark matter $f(\boldsymbol v)$. The second expression determines the normalization of the velocity distribution function. Defining the density fluctuation as
$$
\delta\rho(x) = \rho(x) - \langle \rho(x) \rangle , 
$$
we find the correlation of the density fluctuations as
\begin{align}
\left\langle 
\delta \rho_F(x) 
\delta \rho_F(y) 
\right\rangle
& = \int \frac{d^4k}{(2\pi)^4} e^{i k \cdot (x-y)} {\cal P}^F_{\delta\rho}(k) , 
\\
\left\langle 
\delta \rho_S(x) 
\delta \rho_S(y) 
\right\rangle
& = \int \frac{d^4k}{(2\pi)^4} e^{i k \cdot (x-y)} {\cal P}^S_{\delta\rho}(k) , 
\end{align}
where the correlation between the fast and slow mode vanishes $\langle \delta\rho_F(x) \delta\rho_S(y) \rangle = 0$ and the power spectrum ${\cal P}(k)$ is defined as
\begin{align}
\big\langle 
\widetilde{\delta\rho}(k) 
\widetilde{\delta\rho}^*(k') 
\big\rangle
= (2\pi)^4 \delta^{(4)}(k-k') {\cal P}_{\delta \rho}(k). 
\end{align}
The subscript $F$ and $S$ denote the power spectrum of the fast and slow modes, respectively. An explicit expression for the power spectrum is
\begin{align}
\!\!\!\! {\cal P}^F_{\delta\rho}(k) 
&=
\frac{m^2}{2}
\int_{1,2}
[R_{12}^{F}]^2
f_1
f_2
(2\pi)^4 \delta^{(4)}( k - k_1 - k_2)
\label{fast_den_ps}
\\
\!\!\!\! {\cal P}^S_{\delta\rho}(k)
&= 
m^2 
\int_{1,2}
[R_{12}^{S}]^2 
f_1
f_2
(2\pi)^4 \delta^{(4)}( k - k_1 + k_2)
\label{slow_den_ps}
\end{align}
where $f_{1,2} = f(\boldsymbol v_{1,2})$ and $\int_{1,2} = \int d^3 v_1 d^3v_2$. The above expression for ${\cal P}^F_{\delta \rho}(k)$ is valid for $\omega >0$. The spectrum for $\omega<0$ can be obtained by inverting the frequency $\omega \to -\omega$.  The above power spectra characterize the statistical properties of the ULDM density fluctuation, metric fluctuations, and subsequently, the fluctuations in the angular deflection. 

The power spectrum can be obtained analytically for certain velocity distributions. We assume that the underlying velocity distribution is given by the normal distribution with zero mean, 
\begin{align}
f(v)
= \frac{\bar\rho/m}{(2\pi \sigma_v^2)^{3/2}} 
\exp\left[ - \frac{v^2}{2\sigma_v^2} \right] , 
\end{align}
where $\sigma_v = 160\,{\rm km/sec}$ is the dark matter velocity dispersion in our Galaxy. In this case, we find the power spectrum of the fast and slow density fluctuations as
\begin{align}
\!\!{\cal P}_{\delta\rho}^F(k) 
&= 
(2\pi^2 \bar \rho^2 \lambda^3 \tau) 
\frac{v_k^4}{16} 
\left[ \frac{\bar v^2}{\sigma_v^2} - \frac{v_k^2}{4\sigma_v^2} \right]^{1/2} 
e^{-\bar v^2/\sigma_v^2} , 
\label{fast_density_fluctuation}
\\
\!\!{\cal P}_{\delta\rho}^S(k) 
&= 
(2\pi^2 \bar \rho^2 \lambda^3 \tau) \frac{\sigma_v}{v_k}
\exp\left[ - 
\Big(
\frac{v_k^2}{4\sigma_v^2} + \bar\omega^2 \frac{\sigma_v^2}{v_k^2}
\Big)
\right], 
\label{slow_density_fluctuation}
\end{align}
where $\bar\omega = \omega \tau$, $\bar v^2 = \omega/m -2$, and $\tau = 1 / m \sigma_v^2$. An explicit computation of the two-point function of density fluctuation at the same spacetime point leads to $\langle \delta \rho_S^2(x) \rangle = \bar\rho^2$, and $\langle \delta\rho_F^2(x) \rangle = (15/8) \bar\rho^2 \sigma_v^4$; the fluctuation in the slow mode is as large as its mean value, while typical fluctuations in the fast mode are suppressed by $\sigma_v^2$ compared to those in the slow mode. 

\subsection{Other Components}
We summarize the properties of other components of the energy-momentum tensor for the following discussion. The energy-momentum tensor is given by
\begin{align}
T_{\mu\nu} = 
\partial_\mu \phi \partial_\nu \phi
- \eta_{\mu\nu}
\Big[
\frac{1}{2} \eta^{\rho\sigma} \partial_\rho \phi \partial_\sigma \phi
+ \frac{1}{2} m^2 \phi^2
\Big].
\end{align}
Each component of the energy-momentum tensor is given by
\begin{align}
T_{00}
&=
\rho
=
\frac{1}{2} \left[ \dot \phi^2 + (\nabla\phi)^2 +m^2\phi^2 \right] , 
\\
T_{0i}
&= u_{i} 
= \dot{\phi} \phi_{,i}  , 
\\
T_{kk} 
& = 3 p
= \frac{3}{2} \left[ \dot\phi^2 - \frac{(\nabla\phi)^2}{3} - m^2\phi^2\right] , 
\\
{\cal T}_{ij}
&= \Pi_{ij}
= 
\partial_i \phi \partial_j \phi - \frac{1}{3} \delta_{ij} (\nabla\phi)^2 , 
\end{align}
where the repeated indices are summed, and ${\cal T}_{ij} = T_{ij} - \frac{1}{3} \delta_{ij} T_{kk}$ is the traceless part of the spatial component of the energy-momentum tensor. 

Similar to the density fluctuation, the other components of the energy-momentum tensor can also be decomposed into fast and slow modes. We find, for the fast mode,
\begin{align}
p^F (x)
&=
\frac{m}{2V} \sum_{1,2} 
P_{12}^F
\Big[ 
a_1 a_2 e^{i( k_1 + k_2) \cdot x}
+{\rm h.c.} \Big] , 
\\
u_i^F (x)
&=
\frac{m}{2V} \sum_{1,2} 
U_{12,i}
\Big[ 
a_1 a_2 e^{i( k_1 + k_2) \cdot x}
+{\rm h.c.} \Big] , 
\\
\Pi_{ij}^F (x)
&= 
\frac{m}{2V} \sum_{1,2} 
\pi_{12,ij}
\Big[ 
a_1 a_2 e^{i( k_1 + k_2) \cdot x}
+{\rm h.c.} 
\Big], 
\end{align}
and, for the slow mode,
\begin{align}
p^S (x)
&=
+ \frac{m}{2V} \sum_{1,2} 
P^S_{12}
\Big[ 
a_1 a^*_2 e^{i( k_1 - k_2 ) \cdot x}
+{\rm h.c.} \Big] ,
\\
u_i^S (x)
&= 
- \frac{m}{2V} \sum_{1,2} 
U_{12,i}
\Big[ 
a_1 a^*_2 e^{i( k_1 - k_2 ) \cdot x}
+{\rm h.c.} \Big] ,
\\
\Pi_{ij}^S (x)
&=
- \frac{m}{2V} \sum_{1,2} 
\pi_{12,ij}
\Big[ 
a_1 a^*_2 e^{i( k_1 - k_2 ) \cdot x}
+{\rm h.c.} \Big],
\end{align}
where the coefficients are defined as
\begin{align}
P_{12}^F 
&=
\frac{1}{2m^2}
\Big[
- m^2 
- \omega_1 \omega_2 
+ \frac{\boldsymbol k_1 \cdot\boldsymbol k_2}{3} 
\Big],
\\
P_{12}^S 
& = 
\frac{1}{2m^2}
\Big[
- m^2 
+ \omega_1 \omega_2 
- \frac{\boldsymbol k_1 \cdot\boldsymbol k_2}{3} 
\Big], 
\\
U_{12,i}
&=
\frac{1}{2m^2}
\Big[ 
\omega_1 k_2^i + \omega_2 k_1^i 
\Big], 
\\
\pi_{12,ij}
& = 
\frac{1}{2m^2}
\Big[
- k_1^i k_2^j
- k_2^i k_1^j
+ \frac{2}{3} \delta_{ij} k_1 \cdot k_2 
\Big]. 
\end{align}
The power spectrum of each component of the energy-momentum tensor can be similarly computed by replacing $R_{12}^{F, S}$ in \eqref{fast_den_ps} -- \eqref{slow_den_ps} with relevant coefficients. 

From these coefficients, we can already estimate the typical level of fluctuations in each mode in units of the background density $\bar\rho$. We find, for fast modes,
\begin{align}
\delta p^F / \bar\rho 
& \sim {\cal O}(1) ,
\\
\delta u^F_i / \bar\rho
& \sim {\cal O}(v) ,
\\
\delta \pi^F_{ij} / \bar\rho
& \sim {\cal O}(v^2) ,
\end{align}
and, for slow modes,
\begin{align}
\delta p^S / \bar\rho
& \sim {\cal O}(v^2) ,
\\
\delta u^S_i / \bar\rho
& \sim {\cal O}(v) ,
\\
\delta \pi^S_{ij} / \bar\rho
& \sim {\cal O}(v^2) . 
\end{align}
This power counting will be used to estimate the contribution of each metric fluctuation in the angular deflection. 

\section{Angular Deflection}\label{app:ad_computation}
In this section, we compute the angular deflection in the flat spacetime with the most generic metric fluctuations of a non-expanding universe.\footnote{We assume that the relevant observational time scale is only a few years. During this time scale, the expansion of the universe can be neglected. For the sources that are located at high redshifts, one needs to include the scale factor $a(t)$ in the above parameterization, and this is a trivial generalization of the current discussion.} The metric fluctuations are parameterized as
\begin{align}
ds^2 =& 
- (1 + 2 \Phi) dt^2 
+ 2 {\cal B}_i dx^i dt
\nonumber\\
&
+ \Big[ \big(1 - 2 \Psi) \delta_{ij} + 2 {\cal E}_{ij} \Big] dx^i dx^j . 
\end{align}
We perform scalar-vector-tensor decomposition. The perturbations ${\cal B}_i$ and ${\cal E}_{ij}$ are then parametrized as
\begin{align}
{\cal B}_{i} &= B_{,i} + S_i
\\
{\cal E}_{ij} &= E_{,ij} + \frac{1}{2} (F_{i,j} + F_{j,i}) + \frac{1}{2} h_{ij}
\end{align}
where $S_i$ and $F_i$ are divergenless vectors, satisfying $S_{i,i} = F_{i,i} =0 $ and $h_{ij}$ is a traceless and transverse tensor with $h_{ii} = h_{ij,i} = h_{ij,j}=0$.

Consider the photon four-momentum $k^\mu$. The photon direction is measured by the proper reference frame of the observer. The proper reference frame is constructed with tetrad coordinate basis $e^\mu_{\; a}$ that is normalized as $\eta_{ab} = g_{\mu\nu} e^\mu_{\; a} e^\nu_{\; b}$. The zeroth basis is chosen to be aligned with the four-velocity of the observer, $u^\mu = e^\mu_{\; 0}$. The photon four-momentum can then be expressed in terms of the observed frequency and the direction,
\begin{align}
k^\mu = \omega_{o} u^\mu - \omega_o n_o^i e^{\mu}_{\; i} , 
\end{align}
where the observed frequency and the direction are
\begin{align}
\omega_o 
&= - g_{\mu\nu} k^\mu u^\nu , 
\\
\omega_o n_o^i 
&= - g_{\mu\nu} k^\mu e^\nu_{\; i}.
\end{align}
The measured origin of the photon in the proper reference frame is therefore given by
\begin{align}
n_o^i 
= \frac{g_{\mu\nu} k^\mu e^\nu_{\; i}}{ g_{\mu\nu} k^\mu e^\nu_{\; 0} }. 
\end{align}
We define the angular deflection as the difference in the direction of photon measured by the observer and the source,
\begin{align}
\delta n^i
= n_o^i - n_s^i 
= \frac{g_{\mu\nu} k^\mu e^\nu_{\; i}}{ g_{\mu\nu} k^\mu e^\nu_{\; 0} }\bigg|_{s}^o. 
\label{ad_def}
\end{align}
The subscript $o$ and $s$ denote the quantity evaluated at the observer and the source frame, respectively.

We expand each quantity to the linear order in the metric perturbation,
\begin{align}
g_{\mu\nu}
&= \eta_{\mu\nu} + \delta g_{\mu\nu} , 
\\
k^\mu
&= k^\mu_0 + \delta k^\mu , 
\\
e^\mu_{\; a}
&= \delta^\mu_{\; a} + \delta e^\mu_{\; a},
\end{align}
where the photon four-momentum in the flat spacetime is parameterized as $k^\mu_0 = \omega_0 ( 1, - \boldsymbol n )$.
With this expansion, we find that the angular deflection can be written as
\begin{align}
\delta n^i 
= [ \delta n^i ]_{\rm L}
+ [ \delta n^i ]_{\rm A} , 
\end{align}
where each term is
\begin{align}
[\delta n^i]_{\rm L}
&= - \frac{P^{ij}(n)}{2\omega_0} \int^{\lambda_o}_{\lambda_s} d\lambda' \, 
\delta g_{\mu\nu,j}(\lambda') k_0^\mu k^\nu_0 , 
\\
[\delta n^i]_{\rm A}
&= 
- \frac{ P^{ij}(n) }{ \omega_0 } 
\eta_{\mu\nu} k^\mu_0 
\Big[
(\delta e^\nu_{\; j})_{o}
- (\delta e^\nu_{\; j})_{s}
\Big]. 
\label{app_ab}
\end{align}
Here $\lambda$ is the affine parameter of the photon and $\lambda_{o,s}$ is the affine parameter at the observation and the emission, respectively. Note that we are expanding the tetrad basis with respect to the observer at rest. We also treat the velocity as a small parameter that is sourced only by a small metric perturbation. In the following subsections, we explicitly compute each term. 

\subsection{Lensing Term}
We first consider the lensing term. Using the explicit parameterization of the metric perturbation, we find
\begin{align}
[ \delta n^i ]_L
= 
\omega_0 P^{ij}(n)
\int^{\lambda_o}_{\lambda_s} d\lambda'
\Big[
\Phi + \Psi 
+ n^k {\cal B}_k 
- n^k n^l {\cal E}_{kl}
\Big]_{,j}
\label{app_lensing}
\end{align}
Note that the integration must be performed along the worldline of the photon from the emission point ($\lambda_s$) to the observation point ($\lambda_o$). To the linear order in the metric perturbation, one can replace $\omega_0 d\lambda' \to dt'$.

\subsection{Aberration Term}
To compute the aberration term, we first need to follow the evolution of the tetrad coordinate. The tetrad basis vectors are parallel-transported along the worldline of the observer,
\begin{align}
0 = u^\nu 
\left[
e^{\mu}_{\; a,\nu} 
+ \Gamma^\mu_{\;\;\nu \rho} e^\rho_{\; a}
\right].
\end{align}
Integrating the above equation along the worldline of the observer, we find
\begin{align}
\delta e^0_{\; 0}(\tau) &= - \Phi(\tau) , 
\\
\delta e^i_{\; 0}(\tau) &=
u^i(\tau_0) - {\cal B}_i(\tau) + {\cal B}_i(\tau_0) - \int^\tau_{\tau_0}d\tau' \Phi_{,i} ,
\\
\delta e^0_{\; i}(\tau) &=
u^i(\tau_0) + {\cal B}_i(\tau_0) - \int^\tau_{\tau_0}d\tau' \Phi_{,i} , 
\\
\delta e^i_{\; j}(\tau)
&= 
\omega_{ij}(\tau_0) +\Psi(\tau) \delta_{ij} - {\cal E}_{ij}(\tau)
- \int^\tau_{\tau_0}d\tau' {\cal B}_{[i,j]}.
\end{align}
where $\tau$ is the proper time of the observer,  $\tau_0$ is an arbitrary initial time, $u^i(\tau_0)$ and $\omega_{ij}(\tau_0) = -\omega_{ji}(\tau_0)$ are integration constants, and ${\cal B}_{[i,j]} = (1/2) ( {\cal B}_{i,j} - {\cal B}_{j,i} )$. To determine some of the integration constants, we use the normalization condition for the tetrad basis vectors, $\eta_{ab} = g_{\mu\nu} e^\mu_{\; a} e^\nu_{\; b}$.

Substituting the perturbed tetrad basis $\delta e^\mu_{\; a}$ and the metric perturbation into \eqref{app_ab}, we find
\begin{align}
[ \delta n^i ]_A
= &
P^{ij}(n) 
\left[
\delta e^0_{\; j} + n^k \delta e^k_{\; j}
\right] \big|^o_s
\nonumber\\
= &
P^{ij}(n) 
\bigg[
u^j(\tau_0) 
+ {\cal B}_j(\tau_0)
+ n^k  \omega_{kj}(\tau_0) 
\nonumber\\
&
- n^k {\cal E}_{kj}(\tau) 
 - \int^\tau_{\tau_0}d\tau' \big( \Phi_{,j} + n^k {\cal B}_{[k, j]}  \big) 
\bigg] \bigg|^{o}_{s} . 
\label{app_aberration}
\end{align}
The expression should be understood as the difference of the quantity inside the squared parenthesis evaluated at the observer and the source frame. 

\subsection{Gauge Transformation}
The way we split the lensing and aberration terms is somewhat arbitrary. As a result, the expression $[\delta n^i]_{\rm L}$ and $[\delta n^i]_{\rm A}$ appear gauge-dependent, although the angular deflection \eqref{ad_def} is defined in a manifestly gauge-invariant way. We show below explicitly at the linear order that the sum is independent under the general coordinate transformation. Before that, we briefly review the gauge transformation of each metric perturbation. 

Consider the general coordinate transformation
\begin{align}
x^\mu 
\to \tilde x^\mu (x) 
 = x^\mu + \xi^\mu(x), 
\end{align}
where $\xi^\mu(x)$ is parameterized as
\begin{align}
\xi^\mu = ( \xi^0, \xi_{i} +\zeta_{,i}).
\end{align}
Here $\xi_i$ is the divergenceless vector, satisfying $\xi_{i,i}=0$. Under this gauge transformation, each metric perturbation transforms as
\begin{align}
\Phi &\to \tilde \Phi = \Phi - \dot{\xi}^0 ,
\\
\Psi &\to \tilde \Psi= \Psi ,
\\
B &\to \tilde B = B + \xi^0 -\dot \zeta ,
\\
E &\to \tilde E = E - \zeta ,
\\
S_i &\to \tilde S_i = S_i - \dot{\xi}_i , 
\\
F_i &\to \tilde F_i = F_i -\xi_i ,
\\
h_{ij} &\to \tilde h_{ij} = h_{ij}. 
\end{align}
Gauge-invariant metric perturbations can be constructed as
\begin{align}
\varphi &= \Phi + \frac{d}{dt} (B - \dot{E}) , 
\\
\psi &= \Psi , 
\\
X_i &= S_i - \dot{F}_i  ,
\\
h_{ij} &= h_{ij} .
\end{align}
In addition to the metric perturbations, the constants in the tetrad basis transform as
\begin{align}
u^i &\to \tilde u^i =  
u^i
+ \dot{\zeta}_{,i}
+ \dot{\xi}_i , 
\\
\omega_{ij} &\to \tilde \omega_{ij} =  
\omega_{ij} + \xi_{[i,j]} . 
\end{align}
In the following section, we combine $[\delta n^i]_{\rm L}$ and $[\delta n^i]_{\rm A}$ altogether, and rewrite metric perturbations in terms of their gauge-invariant counterparts to see the sum is invariant under the gauge transformation. 

\subsection{Total Angular Deflection}
Combining the lensing term \eqref{app_lensing} and aberration term \eqref{app_aberration} and rewriting the metric perturbation with its gauge-invariant counterpart, we find
\begin{align}
[\delta n]^i
\approx
[\delta n]^i_{S}
+ [\delta n]^i_{V}
+ [\delta n]^i_{T}
+ [\delta n]^i_{0}
\end{align}
where
\begin{align}
[\delta n]^i_{S}
&= 
P^{ij} (n)
\bigg[
\omega_0 \int^{\lambda_o}_{\lambda_s} d\lambda'  \, (\varphi+\psi)_{,j}
\nonumber\\
&\qquad\qquad
- \int^{\tau_o} d\tau_o' \varphi_{,j} 
+ \int^{\tau_s} d\tau_s' \varphi_{,j}  
\bigg] , 
\nonumber
\\
[\delta n]^i_{V}
&= 
P^{ij} (n)
n^k
\bigg[
\omega_0 \int_{\lambda_s}^{\lambda_o} d\lambda' X_{k,j}
\\
&\qquad\qquad
- \int^{\tau_o} d\tau_o' X_{[k,j]} 
+ \int^{\tau_s} d\tau_s' X_{[k,j]} 
\bigg] , 
\nonumber\\
[\delta n]^i_{T}
&= 
P^{ij} (n)n^k
\bigg[ -\frac{1}{2} h_{jk}\big|^{\lambda_o}_{\lambda_s}
-\frac{ \omega_0 }{2} \int_{\lambda_s}^{\lambda_o}  d\lambda'  n^l h_{kl,j}
\bigg]. 
\nonumber
\end{align}
Here $\tau_o$ ($\tau_s$) is the proper time of the observer (source) at the time of observation (emission). 
The terms with the integral $\int d\lambda'$ are the angular deflection due to the propagation of the photon from the source to an observer. For the scalar and vector contributions, additional terms with the integral $\int d\tau'$ along the worldline of the observer and the source appear due to the perturbation at the position of the observer and the source. 
The term $[\delta n]_0^i$ consists of the metric perturbation and the four-velocity of the observer and the source evaluated at an arbitrary initial time. It is straightforward to check that this term is also gauge-invariant. Since it does not depend on time, it does not affect the analysis in the main text and we do not provide an explicit expression for it. 

For the following discussion, we provide an expression of the angular deflection in the Fourier space,
\begin{widetext}
\begin{align}
\delta n^i
&= 
P^{ij} (n)
\int \frac{d^4k}{(2\pi)^4} U e^{i k \cdot x_o}
\bigg[
- \frac{1}{2} n^k \tilde h_{jk}
- \frac{k_j}{\omega + \boldsymbol n \cdot \boldsymbol k}
\Big(
\tilde \varphi + \tilde \psi + n^k \tilde X_k - \frac{1}{2} n^k n^l \tilde h_{kl}
\Big)
+ \frac{1}{\omega}
(k_j \tilde \varphi + n^k \tilde X_{[k,} k_{j]} )
\bigg],
\label{full}
\end{align}
\end{widetext}
where
\begin{align}
U  = 1 - e^{i \omega d(1 + \boldsymbol k \cdot \boldsymbol n/\omega)}
\end{align}
This expression will be used to estimate the contribution of each term in the presence of underlying ultralight dark matter fluctuations. 

\subsection{ULDM-induced Perturbations}
Each metric perturbation is sourced by underlying fluctuations in the ultralight dark matter halo. The relation between the metric fluctuation and the fluctuation in the energy-momentum tensor is given by the field equation. To the linear order, one finds
\begin{align}
\nabla^2 \psi
&= 
4 \pi G \rho , 
\nonumber
\\
\dot{\psi}_{,i} - \frac{1}{4} \nabla^2 X_i 
&=
4 \pi G u_i , 
\nonumber
\\
\ddot\psi + \frac{1}{3} \nabla^2 (\varphi - \psi)
&= 
4\pi G p, 
\nonumber 
\\
- (\partial_i \partial_j - \frac{\delta_{ij}}{3} \nabla^2 ) (\varphi - \psi)
-  \dot X_{(i,j)} + \frac{\square}{2}  h_{ij}
&= 
8\pi G \Pi_{ij} , 
\nonumber
\end{align}
where $X_{(i,j)} = (1/2) ( X_{i,j} + X_{j,i})$ and $\square h_{ij} = \ddot h_{ij} - \nabla^2 h_{ij}$. To connect the energy-momentum tensor to the metric perturbation, it is convenient to work in the Fourier space. In the Fourier space, each metric fluctuation is then related to each component of the energy-momentum tensor as
\begin{align}
\widetilde \psi (k)
&=
- \frac{4\pi G}{k^2} \widetilde \rho(k) , 
\label{psi_k}
\\
\widetilde \varphi (k)
&=
 - \frac{4\pi G}{k^2} 
\left[ \left( 1 - \frac{3\omega^2}{k^2}\right) \widetilde \rho(k) +3 \widetilde p(k) \right]  , 
\\
\widetilde X_i (k)
&= 
+ \frac{16\pi G}{k^2} P_{ij}(k) \widetilde u_j(k) , 
\\
\widetilde h_{ij}(k)
&= 
+ \frac{16\pi G}{k^2 - \omega^2} \Lambda_{ij,kl}(k) \widetilde \Pi_{kl} , 
\label{h_k}
\end{align}
where the projection tensors are defined as
\begin{align}
P_{ij}(k)
&= \delta_{ij} - \frac{k^i k^j}{k^2} , 
\\
\Lambda_{ij,kl}(k) 
&=
P_{ik}(k) P_{jl}(k) - \frac{1}{2} P_{ij}(k) P_{kl}(k) . 
\end{align}
As the energy-momentum tensor is quadratic in the field $\phi$, it sources not only scalar but also vector and tensor perturbations. Note that, due to the non-relativistic nature and its spectral contents, the induced tensor perturbations are mainly the result of forced oscillation from ULDM fluctuations. 

Using \eqref{psi_k}--\eqref{h_k} and the expressions in Appendix~\ref{app:ULDM_computation}, we find explicit expressions for fast mode metric fluctuation in the non-relativistic limit as
\begin{align}
\widetilde \psi^F(k)
&= 
+ \frac{2 \pi G}{k^2} \frac{m}{V}
\sum_{1,2} 
v_c^2
{\cal O}_F
\\
\widetilde \varphi^F(k)
&= 
- \frac{2\pi G}{k^2} \frac{m}{V}
\sum_{1,2} 
\big[ v_c^2 + v_d^2 - 3 ( \hat v_c \cdot \boldsymbol v_d)^2 \big] 
{\cal O}_F
\\
\widetilde X^F_i(k)
&= 
- \frac{8 \pi G}{k^2} \frac{m}{V}
\sum_{1,2} P_{ij}(k) v_d^j (v_c \cdot v_d) {\cal O}_F
\\
\widetilde h^F_{ij}(k)
&= 
- \frac{2\pi G}{k^2}
\frac{m}{V}
\sum_{1,2} 
v_k^2 \Lambda_{ij,kl} (k) v_d^k v_d^l 
{\cal O}_F
\end{align}
where we define $v_k = k / m$ and 
\bea
{\cal O}_F = a_1 a_2 (2\pi)^4\delta^{(4)} (k-k_1-k_2) . 
\eea 
The above expressions are valid for $\omega >0$. 
For slow mode, we find
\begin{align}
\widetilde \psi^S(k)
&= 
- \frac{4\pi G}{k^2} \frac{m}{V}
\sum_{1,2} 
{\cal O}_S
\\
\widetilde \varphi^S(k)
&= 
- \frac{4\pi G}{k^2} \frac{m}{V}
\sum_{1,2} 
{\cal O}_S
\\
\widetilde X^S_i(k)
&= 
- \frac{16\pi G}{k^2}
\frac{m}{V}
\sum_{1,2} 
P_{ij}(k) v_c^j 
{\cal O}_S
\\
\widetilde h^S_{ij}(k)
&= 
-\frac{16\pi G}{k^2}
\frac{m}{V} 
\sum_{1,2} 
\Lambda_{ij,kl}(k)  v_c^k  v_c^l 
{\cal O}_S
\end{align}
where 
\bea
{\cal O}_S = a_1 a_2^* (2\pi)^4\delta^{(4)} (k - k_1 + k_2) . 
\eea
We have introduced $\boldsymbol v_c = (\boldsymbol v_1 + \boldsymbol v_2)/2$ and $\boldsymbol v_d = (\boldsymbol v_1 - \boldsymbol v_2)/2$.

From the above expressions, we estimate the relative strength of metric fluctuation in units of $4\pi G \bar\rho$ as
\begin{align}
k^2 \widetilde \psi^F &\sim {\cal O}(v^2) ,
\nonumber
\\
k^2 \widetilde \varphi^F &\sim {\cal O}(v^2) ,
\nonumber
\\
k^2 \widetilde X_i^F &\sim {\cal O}(v^3) ,
\nonumber
\\
k^2 \widetilde h_{ij}^F &\sim {\cal O}(v^4) ,
\nonumber
\end{align}
and
\begin{align}
k^2 \widetilde \psi^S &\sim {\cal O}(1) ,
\nonumber
\\
k^2 \widetilde \varphi^S &\sim {\cal O}(1) ,
\nonumber
\\
k^2 \widetilde X_i^S &\sim {\cal O}(v) ,
\nonumber
\\
k^2 \widetilde h_{ij}^S &\sim {\cal O}(v^2) .
\nonumber
\end{align}
Based on this estimation, we can approximate the angular deflection expression in the Fourier space \eqref{full} as
\begin{align}
[\delta n^i]_F
&= 
- P^{ij} (n)
\int \frac{d^4k}{(2\pi)^4} U e^{i k \cdot x_o}
\frac{k_j}{\omega } \tilde \psi_F (k) , 
\label{fast_app}
\\
[\delta n^i]_S
&=
+ P^{ij} (n)
\int \frac{d^4k}{(2\pi)^4} U e^{i k \cdot x_o}
\frac{k_j}{\omega} \tilde \varphi_S (k) .
\label{slow_app}
\end{align}
Note that, for the slow mode, $\widetilde \varphi_S (k) \approx \widetilde \psi_S (k)$. This justifies the expression used in the main text.

\section{Vector Spherical Harmonics Expansion}\label{app:vsh}
The angular deflection can be decomposed into the vector spherical harmonics as
\begin{align}
\delta n^i (t, n) 
= \sum_{\ell m}
\delta n_{E \ell m}(t) Y^E_{\ell m}(n) + \delta n_{B \ell m}(t) Y^B_{\ell m}(n) ,
\end{align}
where the vector spherical harmonics are defined as
\begin{align}
Y^E_{\ell m}(n)
&= \frac{1}{\sqrt{\ell(\ell+1)}} \nabla Y_{\ell m}(n) , 
\\
Y^B_{\ell m}(n)
&= \frac{1}{\sqrt{\ell(\ell+1)}} n\times \nabla Y_{\ell m}(n) . 
\end{align}
The vector spherical harmonics form an orthonormal basis for vector fields on the sky in the sense that
\begin{align}
\int d\Omega_n Y^Q_{\ell m}(n) \cdot [Y^Q_{\ell' m'}(n)]^*
= \delta_{QQ'} \delta_{\ell\ell'} \delta_{mm'}.
\end{align}
From this orthonormality of the vector spherical harmonics, the coefficient can be obtained by
\begin{align}
\delta n_{E \ell m}(t) 
&= \int d\Omega_n \, \delta n(t,n) \cdot [Y^E_{\ell m}(n)]^{*} ,
\\
\delta n_{B \ell m}(t) 
&= \int d\Omega_n \, \delta n(t,n) \cdot [Y^B_{\ell m}(n)]^{*} . 
\end{align}
The two-point correlation of these coefficients can be obtained as 
\begin{align}
\langle \delta n_{Q\ell m}(f, n) \delta_{Q' \ell' m'}^*(f,n') \rangle
= \frac{\delta(f-f') }{2} S(f) C_{Q\ell m Q'\ell' m'}
\end{align}
where $S(f)$ is the same power spectrum for the angular deflection and the correlation coefficient $C_{Q\ell m Q' \ell' m'}$ is given by~\cite{Book:2010pf}
\begin{widetext}
\begin{align}
C_{Q \ell m Q'\ell'm'}
&= 
\int d\Omega_n d\Omega_{n'} 
\Big[ Y^Q_{\ell m}(n) \Big]^*_i \Gamma^{ij}(n,n')
\Big[ Y^{Q'}_{\ell' m'}(n') \Big]_j 
\\
&= 
\frac{1}{\sqrt{\ell(\ell+1)}}
\frac{1}{\sqrt{\ell'(\ell'+1)}}
\int d\Omega_n d\Omega_{n'} 
Y^*_{\ell m}(n)
Y_{\ell' m'}(n')
\beta^{QQ'}(n,n').
\end{align}
\end{widetext}
The $\beta^{QQ'}(n,n')$ function is given by
\begin{align}
\beta^{EE} (n,n')
&= \nabla_i \nabla'_j \Gamma^{ij}(n,n')
\\
\beta^{BB} (n,n')
&= \nabla_q \nabla'_t
\Big[ \eps^{ipq} \eps^{jkt} n_p n'_k \Gamma^{ij}(n,n') \Big]
\\
\beta^{EB}(n,n')
&= \nabla_i \nabla'_p \Big[ \Gamma^{ij} \eps^{jkp} n_k' \Big]
\end{align}
Note that $\nabla$ and $\nabla'$ denote three-dimensional derivatives with respect to $\boldsymbol x$ and $\boldsymbol x'$, respectively. For instance, $\nabla_i n_j = (\delta_{ij} - n_i n_j)$. Among them, only $\beta^{EE}(n,n')$ is non-vanishing. An explicit computation with the ULDM overlap reduction function \eqref{orf} yields
\begin{align}
C_{E\ell m E \ell' m'} = \frac{8\pi}{9} \delta_{\ell \ell'} \delta_{mm'} \delta_{\ell 1}. 
\end{align}
To the leading order, the ULDM signal contains only the dipole component $\ell=1$; all the other component vanishes. 

\section{Simulation}\label{app:simulation}
To investigate the effects of stochastic gravity wave background on the ULDM search, we simulate the angular deflections in the presence of the gravity wave background and white noise. We detail below how we generate synthetic datasets.  

The angular deflection is given by
$$
s_{AI} = r_{AI} + h_{AI}
$$
where $r_{AI}$ and $h_{AI}$ denote the white noise and gravity wave-induced deflection, respectively. Here $I$ indexes the time-series, $s_{AI} = s_A(t_I)$. The noise is generated according to 
\begin{align}
\langle r_{AI} r_{BJ} \rangle = \frac{1}{2} \delta_{AB} \delta_{IJ} \sigma_A^2. 
\end{align}
We assume that the noise is white; the noise power spectral density is related to the above error by $S_A = \sigma_A^2 \Delta t$ with the cadence of the observation $\Delta t$. For simplicity, we assume $\sigma_A^2 = \sigma_r^2$ for all $A$. As we pixelize the sky for the analysis, we rescale the error as $\tilde\sigma_r = \sigma_r \sqrt{N_{\rm pix} / N_\star}$ to reflect the number of stars in each pixel. Then, we draw $r_{AI} \sim {\cal N}(0, \tilde \sigma_r /\sqrt{2})$, where ${\cal N}(\mu,\sigma)$ is the normal distribution with a mean $\mu$ and a standard deviation $\sigma$.

To simulate the correlated gravitational wave noise, we use~\cite{Book:2010pf}
\begin{align}
\delta n_a^i (t)
&=
\sum_{\lambda = +,\times}
\int_{-\infty}^\infty df \, \int d\Omega \, 
e^{-2\pi i f t}
h_\lambda(f,\hat\Omega) 
F^i_\lambda(\hat\Omega, \hat n_a) , 
\end{align}
where the antenna response function is
\begin{align}
F_\lambda^i(\hat\Omega, \hat n_a)
&= 
\frac{1}{2}
\left[
\frac{(\Omega + n)^i n^k n^l}{1+ n\cdot \Omega}
-n^k \delta_{il}
\right] e^\lambda_{kl}(\hat\Omega) , 
\end{align}
with the polarization tensor $e^\lambda_{ij} ( \hat\Omega)$. The strain in the Fourier space follows
\begin{align}
\langle 
h_\lambda(f, \hat \Omega)
h^*_{\lambda'}(f', \hat \Omega')
\rangle
= \frac{\delta_{\lambda\lambda'}}{2}
\frac{\delta^{(2)}(\hat \Omega, \hat \Omega')}{4\pi}
\frac{\delta(f-f')}{2} S_h(f),
\end{align}
where the spectrum is parameterized as
\begin{align}
S_h(f) = \frac{A_{\rm GW}^2}{f} \left( \frac{f}{f_{\rm yr}} \right)^{2 \alpha}. 
\label{psd_gw}
\end{align}
We assume that the strain is a Gaussian random field. The amplitude of the strain $|h_\lambda(f,\hat\Omega)|$ follows the Rayleigh distribution, while the phase follows the uniform distribution. 

For the practical purpose, we discretize the frequency and the solid angle. We  write the angular deflection as
\begin{align}
\delta n_a^i (t)
=
2\Delta f \Delta \Omega
\sum_{\lambda = +,\times}
\sum_{f_i, \Omega_i}
| h_\lambda(f,\hat\Omega)  |
F^i_\lambda(\hat\Omega, \hat n_a)
\nonumber\\
\times
\cos[ 2\pi f t  - \theta_\lambda(f,\hat\Omega) ]
\label{dn_h}
\end{align}
where $h_\lambda(f,\hat\Omega) = |h_{\lambda}(f,\hat\Omega)| \exp[ i \theta_\lambda(f,\hat\Omega)]$. The phase and the amplitude are generated from
\begin{align}
|h_\lambda(f,\hat\Omega)| 
&\sim {\rm Rayleigh}(\sigma_h/\sqrt{2}) 
\\
\theta_\lambda(f,\hat\Omega) 
&\sim {\cal U}(0,2\pi) 
\end{align}
where ${\cal U}(a,b)$ is the uniform distribution in the range of $x\in [a,b)$, ${\rm Rayleigh}(\sigma)$ is the Rayleigh distribution with the probability distribution function $p(x|\sigma) = (x/\sigma) e^{-x^2/2\sigma^2}$, and
\begin{align}
\sigma_h^2(f) = \frac{S_h(f)}{16\pi \Delta f \Delta\Omega} . 
\end{align}

As discussed in the main text, we choose $A_{\rm GW} = 6\times 10^{-15}$ and $\alpha  = -0.15$. We then draw $h(f,\hat\Omega)$ at each frequency $f \in [1/T, 10/T]$ and each pixel $\hat\Omega_i$ ($i=1, \, \cdots,\, 48$). We generate a time series of angular deflection for both benchmarks from $t=0$ to $t = 10\,{\rm yr}$ with $\Delta t = 4\,{\rm week}$ cadence. Since the second benchmark has a smaller cadence, we rescale the single-exposure error for the second case further by $\sigma_r \sqrt{15\,{\rm min}/4\,{\rm week}}$. The time series is equally spaced. The final product \eqref{dn_h} is then summed with the white noise component. 

\section{Optimal Statistics with SGWB}\label{app:os}
In this section, we summarize the signal-to-noise ratio of the injected stochastic gravity waves with each benchmark and the optimal statistic for ULDM when the correlated SGWB noise is present.

\subsection{Signal-to-Noise Ratio for SGWB}
The signal-to-noise ratio for SGWB can be easily estimated with the same expression as discussed in the main text \eqref{SNR}. The cross-correlation expected for SGWB is
\begin{align}
S_{AB}^{\rm GW}(f)
= S_h(f) \Gamma^{\rm GW}_{AB} , 
\end{align}
where the spectrum is given in \eqref{psd_gw}. The overlap reduction function is~\cite{Book:2010pf}
\begin{align}
\Gamma^{\rm GW}_{AB} 
= [\Gamma^{\rm GW}]^{ij}_{ab}
= \alpha(\theta)
\Big[ \hat e^i_{1} \hat e_1^j + \hat e_2^i \hat e_3^j \Big] , 
\end{align}
where $\cos\theta = \hat n_a \cdot \hat n_b$, 
\begin{align}
\alpha(\theta) = 
\frac{7\cos\theta -5}{24} 
- \frac{\sin^4(\theta/2)}{\cos^2(\theta/2)} \ln[\sin(\theta/2)] , 
\end{align}
and the basis for the overlap reduction function is given by
\begin{align}
\hat e_1 = \frac{\hat n_a \times \hat n_b}{\sqrt{1- (\hat n_a \cdot \hat n_b)^2}},
\quad
\hat e_2 = \hat n_a \times \hat e_1
\quad
\hat e_3 = \hat n_b \times \hat e_1. 
\end{align}
With above expressions and \eqref{SNR}, we find
\begin{align}
{\rm SNR}_{\rm GW}
&= 
\bigg[
\frac{ T }{ (\Delta t)^2 }
\sum_{A \neq B} \frac{ [\Gamma^{\rm GW}_{AB}]^2 }{ \sigma_A^2 \sigma_B^2 } 
\int_{f_l}^{f_u} df \, S_{h}^2(f)
\bigg]^{1/2}
\\
&=
\begin{cases}
0.14  \times
\Big( \frac{N_\star}{10^8} \Big)
\Big( \frac{4\,{\rm week}}{\Delta t} \Big)
\Big( \frac{100\mu{\rm as}}{\sigma_r} \Big)^2
\Big( \frac{T}{10\,{\rm yr}} \Big)^{1-2\alpha}
\\
38 \times
\Big( \frac{N_\star}{10^7} \Big)
\Big( \frac{15\,{\rm min}}{\Delta t} \Big)
\Big( \frac{100\mu{\rm as}}{\sigma_r} \Big)^2
\Big( \frac{T}{10\,{\rm yr}} \Big)^{1-2\alpha}
\end{cases}
\nonumber 
\end{align}
where the first and the second line are for the first and the second benchmark, respectively. Similar to the estimate of SNR$_{\rm DM}$, we approximate the overlap reduction function with the solid angle-averaged one, $(\Gamma^{\rm GW}_{AB})^2 \to \langle (\Gamma_{AB}^{\rm GW})^2 \rangle = 2\times 10^{-3}$. In addition, we assume $\sigma_A^2 = \sigma_r^2$ for all $A$. 

The above estimation assumes uniformly distributed background light sources over the entire sky. If only a small fraction of the sky is observed, then the angular average of the overlap reduction function changes. For instance, if $\Delta\Omega \simeq (1.5\,{\rm deg})^2$ is observed, then $\langle (\Gamma_{AB}^{\rm GW})^2 \rangle =1.4\times 10^{-2}$. Assuming no signal is lost during data processing, the SNR$_{\rm GW}$ estimation is increased by $2.7$. 

When computing the signal-to-noise ratio with numerically simulated data, we use the discrete version of the test statistic. Due to the difference in the discrete sum and continuous integral, the final result from numerical simulation could differ by
\begin{align}
\left[ 
\frac{\sum_{i=1}^\infty S_h^2(i f_l) }{\int_{f_l}^{f_u} df S_h^2(f)}
\right]^{\frac12}
= 
\big[ (1-4\alpha) \zeta(2-4\alpha) \big]^{\frac12}
\simeq \sqrt{2}, 
\end{align}
where $\zeta(x)$ is the Riemann zeta function. 

\subsection{Optimal Statistic with Correlated Noise}\label{app:snr_corr}
The optimal test statistic \eqref{ts} is obtained when the null hypothesis contains only uncorrelated noise. If the null hypothesis contains the correlated noise due to SGWB, the form of optimal statistic changes. To find out the form of $\hat Y$ in such a case, we begin with the likelihood function
\begin{align}
{\cal L}
= \frac{1}{\sqrt{\det \Sigma}} 
\exp\left[ 
- \int_{-\infty}^\infty df \, \tilde s^*(f) \Sigma^{-1}(f) \tilde s(f)
\right]
\end{align}
where $\tilde s(f)$ should be understood as a vector of the angular deflection $\tilde s_A(f)$ and $\Sigma(f)$ is the corresponding covariance matrix,
\begin{align}
\Sigma_{AB}(f) 
&=
\left[
S_A(f) \delta_{AB} + S_{AB}^{\rm GW}(f) + S_{AB}^{\rm DM}(f)
\right]
\nonumber\\
&\equiv
\left[
\tilde\Sigma_{AB}(f) + S_{AB}^{\rm DM}(f)
\right]. 
\end{align}
For the following discussion, we write $S_{AB}^{\rm DM}(f) = \Omega \tilde S_{AB}^{\rm DM}(f)$, where $\Omega$ characterizes the strength of ULDM signal. 

In the weak signal limit $\Omega \ll1$, the log-likelihood ratio can be expanded as~\cite{Creighton:2011zz}
\begin{align}
\ln \Lambda
= \ln\left[ 
\frac{{\cal L}({\cal H}_1)}{{\cal L}({\cal H}_0) }
\right]
= \Omega \hat Y
- \frac{\Omega^2}{2} \hat N^2 
+ {\cal O}(\Omega^3),
\end{align}
where the $\hat Y$ and $\hat N^2$ are defined as
\begin{widetext}
\begin{align}
\hat Y
&= \int^{\infty}_{-\infty} df\,
\left[
\tilde s^*(f) 
( \tilde \Sigma^{-1} \tilde S^{\rm DM} \tilde \Sigma^{-1} ) \tilde s(f)
-  \frac{T}{2} \Tr ( \tilde \Sigma^{-1} \tilde S^{\rm DM}) 
\right] , 
\\
\hat N^2
&= 
\int^{\infty}_{-\infty} df\,
\left[
\tilde s^*(f) 
( \tilde \Sigma^{-1} \tilde S^{\rm DM} \tilde \Sigma^{-1} \tilde S^{\rm DM} \tilde \Sigma^{-1} ) \tilde s(f)
-  \frac{T}{2} \Tr ( \tilde \Sigma^{-1}\tilde S^{\rm DM} \tilde \Sigma^{-1}\tilde  S^{\rm DM}) 
\right]. 
\end{align}
\end{widetext}
When the gravity wave background is absent, the $\hat Y$ reduces to the one used in the main text \eqref{ts}. The maximum likelihood estimator for the signal strength is $\hat \Omega_{\rm MLE} = \hat Y / \hat N^2$, and the maximum log-likelihood ratio becomes ${\rm max}_{\Omega} \ln \Lambda = \hat Y^2 /2 \hat N^2$. In the weak signal limit $\Omega \ll1$, $d\ln\Lambda/d\Omega \approx \hat Y$ is known as a locally optimal detection statistic, which has the largest slope compared to any other non-locally optimal statistics and hence the best-performing statistic in this limit~\cite{Kassam_1988}.

With $\hat \rho = \hat Y / \sigma_{\hat Y}$, the signal-to-noise ratio of the locally optimal statistic $\hat Y$ is given by
\begin{align}
\widetilde {\rm SNR}_{\rm DM}
= \langle \hat \rho \rangle
= \left[ T \int_{f_l}^{f_u} df \, \Tr\big(\tilde \Sigma^{-1} \, S^{\rm DM} \, \tilde\Sigma^{-1} \, S^{\rm DM} \big) \right]^{\frac12}. 
\end{align}
If the white noise covariance matrix is proportional to identity matrix $S_{A} \delta_{AB} = S_N \delta_{AB}$ and if we consider only a small part of the sky $\Delta\Omega \ll1$, then the signal-to-noise ratio can be approximated as
\begin{align}
\widetilde {\rm SNR}_{\rm DM}
= \left[ T \int_{f_l}^{f_u} df \, 
\frac{(S_{AB}^{\rm DM})^2/S_N^2}{(1+ N_\star S_h / 12 S_N)^2}
\right]^{\frac12}
\end{align}
Note that ${\rm SNR}_{\rm GW} \sim N_\star S_h / 12 S_N$. That is, in the presence of the correlated SGWB noise, the signal-to-noise ratio for ULDM is parametrically suppressed by
\begin{align}
\widetilde {\rm SNR}_{\rm DM}
\sim \frac{{\rm SNR}_{\rm DM}}{{\rm SNR}_{\rm GW}} 
\end{align}
Given that ${\rm SNR}_{\rm GW} \sim 100$ for the second benchmark, this results in $0.1$ suppression in the projection of $\rho/\rho_0$ in the main figure.

\bibliography{ref}
\bibliographystyle{utphys}
\end{document}